\documentclass[11pt]{article}

\usepackage{textcomp}
\usepackage{pdfsync}
\usepackage{fancyhdr}
\usepackage{amssymb}
\usepackage{amsmath}
\usepackage{mathrsfs} 
\usepackage{graphicx}
\usepackage{latexsym}
\usepackage{appendix}
\usepackage{srcltx}
\def\centeron#1#2{{\setbox0=\hbox{#1}\setbox1=\hbox{#2}\ifdim
   \wd1>\wd0\kern.48\wd1\kern-.48\wd0\fi
   \copy0\kern-.48\wd0\kern-.48\wd1\copy1\ifdim\wd0>\wd1
   \kern.48\wd0\kern-.48\wd1\fi}}

\newcommand{\beq}{\begin{equation}}
\newcommand{\eeq}{\end{equation}}
\newcommand{\bea}{\begin{eqnarray}}
\newcommand{\eea}{\end{eqnarray}}
\newcommand{\ba}{\begin{array}}
\newcommand{\ea}{\end{array}}

\begin{document}


\hskip12cm
\vskip3cm

\begin{center}
 \LARGE \bf  Warped $AdS_3$ black hole in minimal massive gravity with first order formalism
\end{center}

\vskip2cm

\centerline{\Large Soonkeon Nam\footnote{nam@khu.ac.kr}\,,
~~Jong-Dae Park\footnote{jdpark@khu.ac.kr} 
}
\hskip2cm

\begin{quote}
Department of Physics and Research Institute of Basic Science, Kyung
Hee University, Seoul 02447, Republic of Korea$^{1,2}$
\end{quote}

\hskip2cm

\vskip2cm

\centerline{\bf Abstract} 
We obtain a black hole solution of minimal massive gravity theory with Maxwell and electromagnetic Chern-Simons terms using the first order formalism. 
This black hole solution can be translated into the spacelike warped $AdS_3$ black hole solution with some parameters' conditions changing their coordinates system into a Schwarzschild one. 
Applying the Wald formalism to this theory with the first order formalism, we also find out the entropy, mass, and angular momentum of this black hole solution satisfies the first law of black hole thermodynamics. Under the assumption that minimal massive gravity theory with suitable asymptotically warped $AdS_3$ boundary conditions is holographically dual to a two dimensional boundary conformal field theory, we find appropriate central charges by using the relations between entropy of this black hole and the Cardy formula in dual conformal field theory described by left and right moving central charges and temperatures.

\thispagestyle{empty}
\renewcommand{\thefootnote}{\arabic{footnote}}
\setcounter{footnote}{0}
\newpage

\renewcommand{\thesection}{\Roman{section}.}
\renewcommand{\theequation}{\arabic{section}.\arabic{equation}}
\setcounter{section}{0}

\setcounter{equation}{0}
\section{Introduction}
%
There are several three dimensional gravity theories that are suggested to explain various physical problems for example topologically massive gravity (TMG) 
\cite{Deser:1981wh,Deser:1982vy}, new massive gravity (NMG) \cite{Bergshoeff:2009hq}, minimal massive gravity (MMG) \cite{Bergshoeff:2014pca}, etc. 
Various three dimensional gravity theories can be represented by the Chern-Simons-like Lagrangian \cite{Bergshoeff:2014bia}. 
It is well known that an extension of general relativity in three dimensions is TMG, which is composed of the Einstein-Hilbert term, cosmological constant, 
and gravitational Chern-Simons term, breaking parity symmetry. There exists a single massive spin two mode on the linearization of this theory. 
TMG theory also allows some black hole solutions with $AdS_3$ asymptotics \cite{Clement:1992ke,Clement:1994sb,Moussa:2003fc,Moussa:2008sj,Anninos:2008fx}.

In the viewpoint of $AdS/CFT$ correspondence, there exists a discrepancy between any three dimensional gravity theory with asymptotic $AdS_3$ geometry and its dual conformal field theory (CFT) on the boundary. Whenever the spin-two graviton modes propagating on the bulk have positive energy, the central charge of a dual boundary CFT is negative, i.e., non-unitary CFT. This discrepancy is also related to a problem that asymptotic $AdS_3$ black hole solutions have negative mass values whenever bulk graviton modes have positive energy, which is called ``bulk vs boundary clash." An alternative method has been suggested to circumvent this discrepancy, which is ``MMG" theory \cite{Bergshoeff:2014pca}. 
 
One of the purposes of this paper is to construct a black hole solution in MMG theory including Maxwell and electromagnetic Chern-Simons terms with the first order formalism. An intrinsically rotating black hole has been found to TMG theory including Maxwell and electromagnetic Chern-Simons term \cite{Moussa:2008sj}. 
The authors of \cite{Moussa:2008sj} have used a dimensional reduction procedure of \cite{Moussa:1996gm} to search for a stationary rotating black hole solution. In this paper we shall use the first order formalism to find a black hole solution. MMG theory can be represented by a Lagrangian which is composed of that of TMG theory including the torsion term coupled with an auxiliary field $h$ and another ``$ehh$'' term with a dimensionless coupling constant $\alpha$. Auxiliary field $h$ has the same odd-parity and mass dimension with the spin connection $\omega$. The MMG field equation cannot be obtained from an action represented by the metric alone. Elimination of the auxiliary field $h$ from the MMG action cannot be reduced to the action for the metric only, leading to the correct field equation through the variation of the action for the metric. Using linearization of the field equation, it is possible to create parameter regions to evade the bulk vs boundary clash with the condition of positivity of the central charge \cite{Bergshoeff:2014pca}. Therefore, we use the first order formalism to find a black hole solution persisting with the Lagrangian form with auxiliary field $h$. The solution we have found is the spacelike warped $AdS_3$ black hole under the theory we are considering. Similar black hole solutions have been studied in \cite{Moussa:2003fc,Moussa:2008sj,Anninos:2008fx,Nutku:1993eb,Gurses:1994bjn,Setare:2017nlu}. 

Another purpose of this paper is to calculate the mass, angular momentum and entropy of the black hole to be founded under the theory we are considering. Various methods to calculate entropy, mass, and angular momentum as conserved charges have been developed and we have introduced and referred these methods to \cite{Nam:2016pfp}. We use the Wald's formalism to obtain the entropy, mass, and angular momentum of black holes with the first order formalism. In this method we can define charge variation forms to calculate these physical quantities. The entropy of black holes can be obtained by an integration of the charge variation form at bifurcation surface $\mathcal H$, i.e., the event horizon. The mass and angular momentum of black holes can also be obtained by an integration of the charge variation form at spatial infinity, respectively. This method becomes a good working tool in case there are some boundaries at infinity such as asymptotically Minkowski and anti-de Sitter ($AdS$) space-time. Applying this method to the warped $AdS_3$ black hole solution in the theory we are considering, 
we calculate the entropy, mass, and angular momentum of this black hole. These physical quantities are different from those of the warped $AdS_3$ black hole in theTMG case. These values are satisfied with the first law of black hole thermodynamics and are also represented by a Smarr relation. In \cite{Setare:2017nlu} there have been investigations for the warped $AdS_3$ black hole in generalized minimal massive gravity (GMMG). It has been shown that the spacelike warped $AdS_3$ black hole \cite{Tonni:2010gb} is a solution of GMMG which has an additional higher curvature terms occurring in new massive gravity (NMG) \cite{Setare:2014zea}. The authors of \cite{Setare:2017nlu} have calculated the entropy, mass, and angular momentum of this black hole using a different method from ours.

Finally, we have investigated that the entropy of a black hole can be represented by descriptions of quantities of the dual CFT side according to the prescription of $AdS/CFT$ correspondence. By using the Cardy formula, the entropy of a black hole can be described by central charges and temperatures of the dual CFT. If there exists a holographic dual CFT for the MMG theory, then we can find the corresponding central charges by using the Cardy formula with the thermodynamical quantities for the black hole. The warped $AdS_3$ vacua has been referred in \cite{Arvanitakis:2014yja}.

This paper is organized as follows. In Sec. \ref{sec-2}, we survey the Lagrangian and construct equations of motion in MMG theory with Maxwell and electromagnetic Chern-Simons term by using the first order formalism. In Sect. \ref{sec-3}, we find a black hole solution to solve the field equations. In Sect. \ref{sec-4}, we briefly review the Wald formalism to investigate thermodynamic properties of the black hole. Using Wald formalism we define the entropy, mass, and angular momentum of a black hole with the first order formalism, and find these quantities for the black hole solution we have found. Comparing the entropy of the black hole with the Cardy formula we find central charges. In Sec. \ref{sec-5}, we summarize our results and add some comments. Appendices are attached to explain some formulas, and to compute some equations, entropy, mass, and angular momentum of the warped $AdS_3$ black hole in MMG theory.


\setcounter{equation}{0}
\section{Minimal massive gravity theory with Maxwell and electromagnetic Chern-Simons terms}
\label{sec-2}
%
In MMG theory the Lagrangian can be written in terms of the Lorentz vector-valued 1-form $e^a$, dualized connection 
1-form $\omega_a = 1/2 \cdot \epsilon_{abc} \omega^{bc}$ and auxiliary field $h^a$. 
From these the local Lorentz covariant torsion and curvature 2-forms can be defined by 
\begin{equation}\label{def-TC}
       T(\omega) = de + \omega \times e \,, ~~~ R(\omega) = d\omega + \frac{1}{2} \omega \times \omega \,.
\end{equation}
The Lagrangian 3-form for MMG theory is given by
\begin{eqnarray}\label{Lag-mmg}
    {\mathscr L}_{MMG} = - \sigma e \cdot {\cal R} (\omega) + \frac{\Lambda_0}{6} e \cdot e \times e + h \cdot T(\omega)    
              + \frac{1}{2\mu} \Big( \omega \cdot d\omega + \frac{1}{3} \omega \cdot \omega \times \omega \Big) + \frac{\alpha}{2} e \cdot h \times h \,,
\end{eqnarray}
where  $\sigma$ is a sign and $\Lambda_0$ a cosmological constant. Lorentz indices $a, b, c, \cdots$ are suppressed, and operations ``$\cdot$'' and ``$\times$'' represent contractions of $\eta_{ab}$ and $\epsilon_{abc}$ with wedge products. The third term describes the ``Local Lorentz Chern-Simons'' term with a mass 
parameter $\mu$. The fourth term is introduced to avoid the bulk vs boundary clash with a massless parameter $\alpha$. Because the three-dimensional 
Newton constant has inverse mass dimension, the Lagrangian 3-form (\ref{Lag-mmg}) should have mass-squared dimensions. So the cosmological constant 
$\Lambda_0$ has mass squared dimensions. The dreibein $e^a$ can be assigned zero mass dimension and even parity. If we assign the same 
mass dimension and odd parity to auxiliary field $h^a$ and connection $\omega^a$, then the ``Lorentz Chern-Simons" term is the only parity breaking term.
We also include the Maxwell and electromagnetic Chern-Simons terms 
\begin{equation}
   {\mathscr L}_{MC} = - \frac{1}{2} F \wedge * F - \frac{\mu_E}{2} A \wedge dA  \,,
\end{equation}
in the Lagrangian 3-form with a massless dimension parameter $\mu_E$. In order to use the first order formalism, we change the above action with dreibein as follows:
\begin{equation}
    {\mathscr L}_{MC} = - \frac{1}{4} F_{ab} {\mathscr F}_c e^a \wedge e^b \wedge e^c = - \frac{1}{2} F \wedge {\mathscr F} \,,
\end{equation}
where
\begin{equation}\label{em-c-F}
     {\mathscr F} = * F + \mu_E A = \Big( \frac{1}{2} \epsilon_{cfg} F^{fg} + \mu_E A_c \Big) e^c   \,.
\end{equation}
Therefore, the total Lagrangian with the gravitational constant can be represented by
\begin{equation}
    {\mathscr L}_{tot} = {\mathscr L}_{MMG} + {\mathscr L}_{MC}  \,,
\end{equation}
where $\kappa$, i.e., the gravitational constant $8\pi G$, is absorbed in ${\mathscr L}_{MC}$.
To find equations of motion we consider the variation of the total Lagrangian 3-form with respect to $e$, $h$ and $\omega$. Then the variation for ${\mathscr L}_{tot}$ 
is given by
\begin{eqnarray}
   \delta {\mathscr L}_{tot} &=& \delta e \cdot \Big( - \sigma R(\omega) + \frac{\Lambda_0}{2} e \times e + D(\omega) h 
                     + \frac{\alpha}{2} h \times h + \mathcal{T} \Big)       \nonumber   \\
               && + \delta \omega \cdot \Big( - \sigma T(\omega) + e \times h + \frac{1}{\mu} R(\omega) \Big)            \nonumber   \\
               && + \delta h \cdot \Big( T(\omega) + \alpha e \times h \Big) -  \delta A \wedge d{\mathscr F}   \nonumber   \\
               && + d \Big( \delta e \cdot h - \sigma \delta \omega \cdot e + \frac{1}{2\mu} \delta \omega \cdot \omega - \delta A \wedge {\mathcal F} \Big)    \,,   \label{var-Lag-t}
\end{eqnarray}
where
\begin{equation}\label{em-eq-F}
      {\mathcal F} = * F + \frac{\mu_E}{2} A \,.
\end{equation}
So, we can get equations of motion as follows:
\begin{eqnarray}
     T(\omega) + \alpha e \times h  = 0  \,,     \label{o-eq-1} \\
     R(\omega) + \mu e \times h - \sigma \mu T(\omega)  = 0 \,,     \label{o-eq-2}  \\
     -\sigma R(\omega) + \frac{\Lambda_0}{2} e \times e + D(\omega) h + \frac{\alpha}{2} h \times h + {\mathcal T} = 0 \,,    \label{o-eq-3}
\end{eqnarray}
where the 2-form field ${\mathcal T}_a$ associated with the matter part is given by
\begin{equation}\label{matt-eq}
   {\mathcal T}_a = - \frac{1}{4} \big( 2 F_{ab} {\mathscr F}_c + {\mathscr F}_a F_{bc} \big) e^b \wedge e^c \,.
\end{equation}
Now we can use (\ref{o-eq-1}) to make the torsion free condition
\begin{equation}
    T(\Omega) = de + \Omega \times e = 0  \,, 
\end{equation}
then we should consider the shifted connection $\Omega = \omega + \alpha h$. Using these new connection 1-forms we can reexpress equations of motion as follows:
\begin{eqnarray}
     T(\Omega) = 0 \,,     \label{cmg-eq-1} \\
     R(\Omega) + \frac{\alpha \Lambda_0}{2} e \times e  + \mu (1+\sigma\alpha)^2 e \times h  
            + \alpha {\mathcal T} = 0 \,,     \label{cmg-eq-2}  \\
     D(\Omega) h - \frac{\alpha}{2} h \times h + \sigma \mu (1+\sigma\alpha) e \times h    
          + \frac{\Lambda_0}{2} e \times e + {\mathcal T} = 0 \,.     \nonumber  \\      \label{cmg-eq-3}
\end{eqnarray}
From the variation with respect to the gauge field $A$, The equation of motion for the Maxwell and electromagnetic Chern-Simons terms is given by
\begin{equation}\label{em-eq}
    d{\mathscr F} = d \big( * F + \mu_E A \big) =  0  \,.
\end{equation}
%

\setcounter{equation}{0}
\section{Black hole solution}
\label{sec-3}
%

%
To find a black hole solution with stationary circular symmetry, the dimensional reduction method has been used  in \cite{Clement:1992ke,Clement:1994sb,Moussa:2003fc,Moussa:2008sj,Clement:2009gq}. Following the procedure of this method, we can take a metric ansatz with two commuting Killing vectors 
$\partial_t$ and $\partial_{\phi}$ as 
\begin{equation}\label{metric-1}
     ds^2 = \frac{d\rho^2}{\zeta(\rho)^2 f(\rho)^2} + \lambda_{\mu\nu} (\rho) dx^{\mu} dx^{\nu} \,,
\end{equation}
where $x^{\mu}$ expresses two coordinates $t, \phi$ and $f(\rho)^2 = - \det \lambda$. The function $\zeta(\rho)$ is introduced as a scale factor for arbitrary 
reparametrizations of the coordinate $\rho$. If we assume that the space-time has two commuting Killing vectors $\partial_t$ and $\partial_{\phi}$, 
the special linear group $SL(2, \mathbb{R})$ of transformations in the Killing vector space is locally isomorphic to the Lorentz group $SO(1, 2)$. Therefore the parametrization 
of the matrix $\lambda_{\mu\nu}$ can be described by
\begin{eqnarray}
     \lambda = \left( \begin{array}{cc}
                                   T(\rho) + X(\rho)  &  Y(\rho)     \\
                                   Y(\rho)    &   T(\rho) - X(\rho)
                             \end{array}  \right)  \,.
\end{eqnarray}
The special linear transformation of $\lambda_{ab}$ corresponds to the Lorentz transformation of a vector $\vec{X} = (T, X, Y)$, and 
$f(\rho)^2 = - \det \lambda = \vec{X}^2 = - T^2 + X^2 + Y^2$ is the pseudo-norm in Minkowski space. We now take 
\begin{equation}
     T(\rho) - X(\rho) = R(\rho)^2 \,, ~~~ Y(\rho) = h(\rho) \,,
\end{equation}
then (\ref{metric-1}) becomes
\begin{equation}\label{met-ans}
     ds^2 = - \frac{f(\rho)^2}{R(\rho)^2} dt^2 + \frac{d{\rho}^2}{\zeta(\rho)^2 f(\rho)^2} + R(\rho)^2 \bigg( d\phi + \frac{h(\rho)}{R(\rho)^2} dt \bigg)^2  \,.
\end{equation}
For the above metric, we can obtain dreibein as follows:
\begin{equation}\label{drei}
    e^0 = \frac{f}{R} dt \,, ~~~ e^1 = \frac{1}{\zeta f} d\rho \,, ~~~ e^2 = R \bigg( d\phi + \frac{h}{R^2} dt \bigg) \,,
\end{equation}
where we simply omit the representation of coordinate $\rho$ from functions of $\rho$ such as $f$, $\zeta$, $R$ and $h$.
From now on, using this dreibein we will develop the first order formalism instead of the dimensional reduction method. 
To solve the equations of motion we should find the components of the electro-magnetic field $F_{ab} = e^{\mu}_{~a}  e^{\nu}_{~b} F_{\mu\nu}$ which can be represented by
\begin{equation}\label{em-F-1}
    F_{01} = - \zeta R \bigg( A'_t - \frac{h}{R^2} A'_{\phi} \bigg) \,, ~~~ F_{12} = \zeta \frac{f}{R} A'_{\phi} \,,
\end{equation}
where ` $\prime$ ' means the differentiation with respect to $\rho$, and other components vanish. Also, we can get the components of (\ref{em-c-F})
\begin{eqnarray}\label{em-F-2}
   &&  {\mathscr F}_0 = \zeta \frac{f}{R} A'_{\phi} + \mu_E \frac{R}{f} \Big( A_t - \frac{h}{R^2} A_{\phi} \Big)  \,,     \nonumber   \\
   &&  {\mathscr F}_1 = 0   \,,        \\
   &&  {\mathscr F}_2 = \zeta R \Big( A'_t - \frac{h}{R^2} A'_{\phi} \Big) + \mu_E \frac{1}{R} A_{\phi}  \,.   \nonumber 
\end{eqnarray}
Using dreibein (\ref{drei}) and the above components of the electromagnetic field we firstly consider the equation of motion for the electromagnetic field (\ref{em-eq}) to find a solution. Then we can obtain two equations
\begin{eqnarray}
     \zeta R \Big( \frac{f}{R} {\mathscr F}_0 \Big)' + {\mathscr F}_2 \zeta R^2 \Big( \frac{h}{R^2} \Big)'  = 0 \,,     \\
     \zeta \frac{f}{R} ({\mathscr F}_2 R)' = 0 \,.  
\end{eqnarray}
These equations can be easily solved as 
\begin{equation}\label{em-F-3}
    {\mathscr F}_2 = \frac{C_2}{R} \,, ~~~ \frac{f}{R} {\mathscr F}_0 + C_2 \frac{h}{R^2} = C_0  \,.
\end{equation}
For the simplest case we consider constants $C_0=C_2=0$. Then we can find ${\mathscr F}_a = 0$.  
Applying these results to (\ref{em-F-2}) we get
\begin{eqnarray}
     \zeta R^2 \Big( A'_t - \frac{h}{R^2} A'_{\phi} \Big) + \mu_E A_{\phi} = 0  \,,  \label{em-diff-1} \\
     \zeta \frac{f^2}{R^2} A'_{\phi} + \mu_E \Big( A_t - \frac{h}{R^2} A_{\phi} \Big) = 0  \,.   \label{em-diff-2}
\end{eqnarray} 
If we take $\zeta = \mu_E$, then the second equation can be simply reduced to  
\begin{equation}\label{em-diff-3}
     f^2 A'_{\phi} + R^2 A_t - h A_{\phi} = 0 \,.     
\end{equation}
To solve the above equation let us take functions as some polynomials 
\begin{eqnarray}
   f^2 &=& f_2 {\rho}^2 + f_1 \rho + f_0 \,, \nonumber  \\  
   h&=&h_1 \rho + h_0 \,,        \nonumber  \\
   R^2 &=& c_2 {\rho}^2 + c_1 \rho + c_0 \,,    \label{funcs-1}      \\   
   A_t &=&T_1 \rho + T_0 \,, \nonumber  \\ 
   A_{\phi}&=& \phi_1 \rho + \phi_0 \,.     \nonumber
\end{eqnarray}
Substituting the above functions into (\ref{em-diff-3}), then we can find an equation of $\rho$ which has cubic term as its maximum order. If we require 
the cubic term to be vanished, then we can choose $T_1=0$ which means that the electric field does not make any physical effect. The other coefficients 
of $\rho$ terms give us three relations as follows:
\begin{eqnarray}
    f_2 \phi_1 + c_2 T_0 -h_1 \phi_1 = 0  \,,   \nonumber  \\
    f_1 \phi_1 + c_1 T_0  - h_1 \phi_0 - h_0 \phi_1 = 0  \,,    \label{rels}   \\
    f_0 \phi_1 + c_0 T_0 - h_0 \phi_0 = 0 \,,     \nonumber 
\end{eqnarray}
Substituting functions (\ref{funcs-1}) into (\ref{em-diff-1}) with $\zeta = \mu_E$, we can get
\begin{equation}\label{funcs-2}
     h_1 = 1 \,, \quad  \phi_0 = h_0 \phi_1 \,, \quad  T_0 = \frac{\phi_1}{c_2} (1 - f_2) \,.
\end{equation}
With these values we can solve the equations in (\ref{rels}) to obtain two coefficients
\begin{equation}\label{c-conds}
     c_1 = \frac{c_2 (2h_0 - f_1)}{1 - f_2} \,, \qquad  c_0 = \frac{c_2 (h_0^2 - f_0)}{1 - f_2} \,.
\end{equation}
If $f_2 \neq 0$, then the linear term of function $f^2 = f_2 \rho^2 + f_1 \rho + f_0$ can be set to zero $(f_1 = 0)$ by a translation of $\rho$.
Therefore, we can simply represents three functions in the metric ansatz (\ref{met-ans}) as follows:
\begin{eqnarray}\label{sol-func}
     && R(\rho)^2 = \frac{c_2}{1 - f_2} \big( h^2 - f^2 \big)  \,,   \nonumber  \\
     && f(\rho)^2 = f_2 (\rho^2 - \rho_0^2)  \,,        \\  
     && h(\rho) = \rho + h_0 \,.   \nonumber
\end{eqnarray}
We now investigate the equations of motion, (\ref{cmg-eq-1}), (\ref{cmg-eq-2}), and (\ref{cmg-eq-3}), to find a black hole solution. First, we consider the torsion free condition (\ref{cmg-eq-1}) for the metric ansatz (\ref{met-ans}). It is convenient to use some relations derived from (\ref{sol-func}) while we calculate 
connection 1-forms, curvature 2-forms, and auxiliary fields:
\begin{eqnarray}
       && \Big( \frac{f^2}{R^2} \Big)' = \Big( \frac{h^2}{R^2} \Big)'  \,, \nonumber  \\ 
       && \Big( \frac{f}{R} \Big)' = \frac{h}{f R} \bigg( 1- \frac{1}{2} h \frac{(R^2)'}{R^2} \bigg) \,,    \\
       && \Big( \frac{h}{R} \Big)' = \frac{1}{R} \bigg( 1 - \frac{1}{2} h \frac{(R^2)'}{R^2} \bigg)  \,.  \nonumber  
\end{eqnarray}
By substituting (\ref{drei}) into (\ref{cmg-eq-1}) we can find connection 1-forms 
\begin{eqnarray}
    && \Omega^0 = \frac{1}{2} \zeta \bigg( 1 - h \frac{(R^2)'}{R^2} \bigg) e^0 + \frac{1}{2} \zeta f \frac{(R^2)'}{R^2}  e^2 \,,       \nonumber    \\
    && \Omega^1 = \frac{1}{2} \zeta \bigg( 1 - h \frac{(R^2)'}{R^2} \bigg) e^1 \,,      \\
    && \Omega^2 = - \frac{1}{2} \zeta \bigg( 1 - h \frac{(R^2)'}{R^2} \bigg) e^2 + \zeta \frac{h}{f} \bigg( 1- \frac{1}{2} h \frac{(R^2)'}{R^2} \bigg) e^0  \,,     \nonumber 
\end{eqnarray}
where we regard $\zeta$ as a constant because we take $\zeta = \mu_E$ before. We can also find the curvature 2-forms from the definition (\ref{def-TC}) 
with the above shifted connection 1-forms $\Omega^a$,
\begin{eqnarray}
    && R^0 = \bigg\{ \frac{1}{4} \zeta^2 + \frac{1}{2} \zeta^2 (R^2)''  \Big( \frac{f^2}{R^2} \Big)  \bigg\} e^1 \wedge e^2   
                       - \frac{1}{2} \zeta^2 (R^2)''  \frac{f h}{R^2} e^1 \wedge e^0  \,,        \nonumber      \\
    && R^1 = - \frac{1}{4} \zeta^2 e^2 \wedge e^0  \,,    \\
    && R^2 = \bigg\{ - \frac{1}{4} \zeta^2 + \frac{1}{2} \zeta^2  (R^2)'' \Big( \frac{h^2}{R^2} \Big) \bigg\} e^0 \wedge e^1  
                       + \frac{1}{2} \zeta^2 (R^2)'' \frac{f h}{R^2} e^1 \wedge e^2  \,.         \nonumber   
\end{eqnarray}
We are dealing with a simplest case (\ref{em-F-3}) with $C_0 = C_2 = 0$, i.e., ${\mathscr F}_a = 0$. It means that the matter part of the equations of motion (\ref{cmg-eq-2}) 
and (\ref{cmg-eq-3}) vanished, ${\mathcal T}_a = 0$.  So, we can find auxiliary fields $h^a$ from (\ref{cmg-eq-2}) as follows: 
\begin{eqnarray}
    && h^0 = \frac{1}{\tilde{\mu}} \bigg\{ \frac{\xi^2}{8} - \frac{1}{4} \zeta^2 (R^2)''  \Big( \frac{h^2 + f^2}{R^2} \Big) \bigg\} e^0     
                     + \frac{1}{\tilde{\mu}} \bigg\{ \frac{1}{2} \zeta^2 (R^2)''  \frac{f h}{R^2} \bigg\} e^2 \,,           \nonumber      \\
    && h^1 = \frac{1}{\tilde{\mu}} \bigg\{ \frac{\xi^2}{8}  - \frac{1}{4} \zeta^2 (R^2)''  \Big( \frac{h^2-f^2}{R^2} \Big) \bigg\} e^1  \,,      \\
    && h^2 = \frac{1}{\tilde{\mu}} \bigg\{ \frac{\xi^2}{8} + \frac{1}{4} \zeta^2 (R^2)''  \Big( \frac{h^2 + f^2}{R^2} \Big) \bigg\} e^2      
                     - \frac{1}{\tilde{\mu}} \bigg\{ \frac{1}{2} \zeta^2 (R^2)''  \frac{f h}{R^2} \bigg\} e^0  \,,            \nonumber   
\end{eqnarray}
where parameters have been changed by $\xi^2=\zeta^2 - 4 \alpha\Lambda_0$ and $\tilde{\mu} = \mu(1+\sigma\alpha)^2$.
Now we reexpress Eq.(\ref{cmg-eq-3}) as component forms 
\begin{equation}\label{ch-eq}
    dh_a + \epsilon_{abc} \Omega^b \wedge h^c - \alpha \epsilon_{abc} h^b \wedge h^c      
          + \sigma \mu (1+\sigma\alpha) \epsilon_{abc} e^b \wedge h^c + \frac{\Lambda_0}{2} \epsilon_{abc} e^b \wedge e^c = 0 \,.
\end{equation}
By using the results of dreibein, shifted connections, and auxiliary fields we can find five equations but two of these are the same one. So, the equations resulting from (\ref{ch-eq}) are given by
\begin{eqnarray}
    - \frac{1}{4} \zeta^2 (R^2)'' \zeta \frac{h^2 + 2f^2}{R^2} - \frac{\alpha}{\tilde{\mu}} \bigg( \frac{1}{8} \xi^2 - \frac{1}{4} \zeta^2 (R^2)'' \frac{h^2 - f^2}{R^2} \bigg)
          \cdot \bigg( \frac{1}{8} \xi^2 + \frac{1}{4} \zeta^2 (R^2)'' \frac{h^2 + f^2}{R^2} \bigg)           \nonumber   \\
          + \sigma \mu (1+\sigma\alpha) \bigg( \frac{1}{4} \xi^2 + \frac{1}{2} \zeta^2 (R^2)'' \frac{f^2}{R^2} \bigg) + \tilde{\mu} \Lambda_0 = 0 \,.    \label{ff-eq-1}  \\
   - \frac{1}{4} \zeta^2 (R^2)'' \cdot 3 \zeta \frac{f h}{R^2} 
        - \frac{\alpha}{\tilde{\mu}} \cdot \frac{1}{2} \zeta^2 (R^2)'' \frac{f h}{R^2} \bigg( \frac{1}{8} \xi^2 - \frac{1}{4} \zeta^2 (R^2)'' \frac{h^2 - f^2}{R^2} \bigg) 
        \nonumber    \\
        + \sigma \mu (1+\sigma\alpha) \cdot \frac{1}{2} \zeta^2 (R^2)'' \frac{f h}{R^2} = 0 \,.     \label{ff-eq-2}
   \end{eqnarray}
\begin{eqnarray}
       - \frac{1}{4} \zeta^2 (R^2)'' \zeta \frac{h^2 - f^2}{R^2} - \frac{\alpha}{\tilde{\mu}} \bigg( \frac{1}{8} \xi^2 - \frac{1}{4} \zeta^2 (R^2)'' \frac{h^2 - f^2}{R^2} \bigg) 
                  \cdot \bigg( \frac{1}{8} \xi^2 + \frac{1}{4} \zeta^2 (R^2)'' \frac{h^2 - f^2}{R^2} \bigg)       \nonumber    \\
      + \sigma \mu (1+\sigma\alpha) \frac{1}{4} \xi^2 + \tilde{\mu} \Lambda_0 = 0 \,.                 \label{ff-eq-3}      \\
    \frac{1}{4} \zeta^2 (R^2)'' \zeta \frac{2 h^2 + f^2}{R^2} - \frac{\alpha}{\tilde{\mu}} \bigg( \frac{1}{8} \xi^2 - \frac{1}{4} \zeta^2 (R^2)'' \frac{h^2 - f^2}{R^2} \bigg)
        \cdot \bigg( \frac{1}{8} \xi^2 - \frac{1}{4} \zeta^2 (R^2)'' \frac{h^2 + f^2}{R^2} \bigg)       \nonumber   \\
        + \sigma \mu (1+\sigma\alpha) \bigg( \frac{1}{4} \xi^2 - \frac{1}{2} \zeta^2 (R^2)'' \frac{h^2}{R^2} \bigg) + \tilde{\mu} \Lambda_0 = 0 \,.      \label{ff-eq-4}
\end{eqnarray}
Adding (\ref{ff-eq-1}) to (\ref{ff-eq-4}) and then multiplying $2/\mu\zeta^2$, we can obtain
\begin{equation}\label{condi-1}
    \bigg\{ 2 \sigma (1+\sigma\alpha) - \frac{\zeta}{\mu} - \frac{\alpha \xi^2}{4 \mu \tilde{\mu}} \bigg\} (1-f_2)    
         = \sigma (1+\sigma\alpha) \frac{\xi^2}{\zeta^2} - \frac{\alpha \xi^4}{16 \mu \tilde{\mu} \zeta^2} + 4 (1+\sigma\alpha)^2 \frac{\Lambda_0}{\zeta^2}  \,,
\end{equation}
where we use the first relation for functions in (\ref{sol-func}). Therefore, we can find a constant
\begin{equation}\label{condi-1-1}
   f_2 = \frac{\sigma(1+\sigma\alpha)\Big( 1+\frac{4\alpha\Lambda_0}{\zeta^2} \Big) - \frac{\zeta}{\mu} 
         - \frac{\alpha}{4} \frac{(\zeta^2 - 4\alpha \Lambda_0)}{\mu^2(1+\sigma\alpha)^2}\Big( \frac{3}{4} + \frac{\alpha \Lambda_0}{\zeta^2} \Big)
         - 4(1+\sigma\alpha)^2 \frac{\Lambda_0}{\zeta^2}}{2\sigma(1+\sigma\alpha) - \frac{\zeta}{\mu} 
         - \frac{\alpha}{4} \frac{(\zeta^2 - 4\alpha \Lambda_0)}{\mu^2(1+\sigma\alpha)^2}}  \,.
\end{equation}
By using (\ref{sol-func}) and multiplying $\xi^2/4 \zeta^2$, Eq.(\ref{ff-eq-2}) becomes 
\begin{equation}\label{condi-3}
   \frac{\alpha \xi^2}{4 \mu \tilde{\mu}} (1-f_2) = - \sigma(1+\sigma\alpha) \frac{\xi^2}{2 \zeta^2} + \frac{3\xi^2}{4 \mu \zeta} 
         + \frac{\alpha \xi^4}{16 \mu \tilde{\mu} \zeta^2}  \,. 
\end{equation}
By using (\ref{sol-func}), (\ref{condi-3}), and mutiplying $2/\mu\zeta^2$, Eq. (\ref{ff-eq-3}) becomes
\begin{equation}
     \bigg( 2\sigma(1+\sigma\alpha) - \frac{\zeta}{\mu} \bigg) (1-f_2)   
           = \sigma (1+\sigma\alpha) \frac{\xi^2}{2\zeta^2} + \frac{3\xi^2}{4 \mu \zeta}
            + 4(1+\sigma\alpha)^2 \frac{\Lambda_0}{\zeta^2}  \,.
\end{equation}
Finally, we can obtain
\begin{equation}\label{condi-5}
    f_2 = \frac{\frac{3}{2} \sigma(1+\sigma\alpha) - \frac{7\zeta}{4\mu} + 3 \alpha \frac{\Lambda_0}{\mu \zeta} 
              - 2(1+\sigma\alpha)(2+\sigma\alpha) \frac{\Lambda_0}{\zeta^2}}{2 \sigma(1+\sigma\alpha) - \frac{\zeta}{\mu}}   \,.
\end{equation}
Substituting (\ref{condi-3}) into (\ref{condi-1-1}) we can also find the same result with (\ref{condi-5}).
When $\alpha$ goes to zero, then (\ref{condi-5}) approaches
\begin{equation}\label{zm1}
    f_2 \longrightarrow \frac{\sigma - \frac{\zeta}{\mu} - \frac{4\Lambda_0}{\zeta^2}}{2\sigma - \frac{\zeta}{\mu}}  \,, 
\end{equation}
and (\ref{condi-3}) approaches
\begin{equation}\label{zm2}
         \frac{3\zeta}{4\mu} \rightarrow \frac{\sigma}{2} \,.
\end{equation}
These results can be identified with the same results (3.9) and (3.13) in \cite{Moussa:2008sj}. Let $c_2 =1$ and $f_2 = \beta^2$ for convenience, 
and set $f_0 = - \beta^2 \rho_0^2$ and $h_0 = \omega (1-\beta^2)$. Then we can represent three functions (\ref{sol-func}) as follows:
\begin{eqnarray}
     && R^2 = \rho^2 + 2 \omega \rho + \omega^2 (1 - \beta^2) + \frac{\beta^2 \rho_0^2}{1 - \beta^2}  \,,      \nonumber   \\
     && f^2 = \beta^2 (\rho^2 - \rho_0^2)  \,,   \\
     && h = \rho + \omega (1 - \beta^2)  \,.       \nonumber    
\end{eqnarray}
By using these functions  we can rewrite this black hole solution
\begin{equation}
    ds^2 = - \frac{\beta^2 (\rho^2 - \rho_0^2)}{R^2} dt^2 + \frac{d\rho^2}{\zeta^2 \beta^2 (\rho^2 - \rho_0^2)}    
            + R^2 \bigg( d\phi + \frac{\rho + \omega(1-\beta^2)}{R^2} dt \bigg)^2  \,,
\end{equation} 
which is the same form of \cite{Moussa:2008sj} and \cite{Clement:2009gq} with a parameter condition $0 < \beta^2 < 1$ for causally regular black hole solution.
%
Taking $\sigma = 1$ for convenience and solving this inequality, we can find ranges of parameters $\alpha$, $\Lambda_0$ and $\zeta/\mu$.
The allowed ranges for parameters are represented by Table \ref{table1}.

%
\begin{table}[h]
\caption{\label{table1} A table of parameters ranges for causally regular black hole. Ranges \textsf{A} and \textsf{B} for $\zeta/\mu$ are represented by Eq. (\ref{FA})
and Eq. (\ref{FB}).}
\begin{center}
\begin{tabular}[t]{ | c | c | c | }
\hline 
   $ \alpha $   &  $\Lambda_0$   & $\zeta/\mu$   \\
 \hline \hline
    $-1 < \alpha < 0 $  &   $ \frac{\zeta^2}{4\alpha} < \Lambda_0 < \frac{\zeta^2}{2(\alpha-1)} $   &   $ {\mathsf B} < \frac{\zeta}{\mu} < {\mathsf A} $   \\
 \hline  
   $ 1 < \alpha < 7 $   & $ \Lambda_0 < \frac{\zeta^2}{4\alpha}$   & $ {\mathsf A} < \frac{\zeta}{\mu} < {\mathsf B} $  \\
        &  $\frac{\zeta^2}{4\alpha} < \Lambda_0 < \frac{7\zeta^2}{12\alpha}$   & $ \frac{\zeta}{\mu} < {\mathsf B}$\,,  or   ${\mathsf A} < \frac{\zeta}{\mu}$   \\
        & $\Lambda_0 > \frac{\zeta^2}{2(\alpha - 1)}$   &   ${\mathsf B} < \frac{\zeta}{\mu} <{\mathsf A}$   \\
\hline 
  $ \alpha > 7 $    &   $\Lambda_0 < \frac{\zeta^2}{4\alpha}$     &    ${\mathsf A} < \frac{\zeta}{\mu} < {\mathsf B}$    \\
       &   $\frac{\zeta^2}{4\alpha} < \Lambda_0 < \frac{\zeta^2}{2(\alpha - 1)}$    &     $\frac{\zeta}{\mu} < {\mathsf B}$  or  ${\mathsf A} < \frac{\zeta}{\mu} $   \\
       &   $\frac{\zeta^2}{2(\alpha - 1)} < \Lambda_0 < \frac{7\zeta^2}{12\alpha}$    &     $\frac{\zeta}{\mu} < {\mathsf A} $ or ${\mathsf B} < \frac{\zeta}{\mu}$    \\
\hline
\end{tabular}
\end{center}
\end{table}
In this Table \ref{table1}, we express \textsf{A} and \textsf{B} as upper and lower limits of the allowed ranges for $\zeta/\mu$ as follows
\begin{eqnarray}
   &&{\mathsf A} = \frac{2(1+\alpha)\Big( (\alpha+2)\frac{\Lambda_0}{\zeta^2} + \frac{1}{4} \Big)}{3\alpha\frac{\Lambda_0}{\zeta^2} - \frac{3}{4}}  \,,  \label{FA} \\
    &&{\mathsf B} = \frac{2(1+\alpha)\Big( (\alpha+2)\frac{\Lambda_0}{\zeta^2} - \frac{3}{4} \Big)}{3\alpha\frac{\Lambda_0}{\zeta^2} - \frac{7}{4}}  \,.   \label{FB}
\end{eqnarray}
%
%

%
If we change the coordinate system and take some parameters as follows
\begin{eqnarray}
    && \rho \rightarrow r - \frac{r_+ + r_-}{2} \,, \quad t \rightarrow \bigg( \frac{4}{3(\nu^2-1)} \bigg)^{1/2} \nu \ell t \,,    \nonumber      \\
    && \phi \rightarrow \bigg( \frac{3(\nu^2-1)}{4} \bigg)^{1/2} \ell \theta \,,  ~~~ \rho_0 \rightarrow \frac{1}{2} (r_+ -  r_-) \,,    \label{beta-zeta}   \\
    && \omega(1-\beta^2) = \frac{1}{2} \big( r_+ + r_- - 2\beta \sqrt{r_+ r_-} \big) \,, ~~ \beta^2 = \frac{\nu^2 + 3}{4\nu^2} \,, ~~  \nonumber      \\
    && \zeta^2 = \frac{4\nu^2}{\ell^2} \,,   ~~ R(\rho)^2 = \frac{4}{3(\nu^2 - 1)} R(r)^2  \,,     \nonumber 
\end{eqnarray}
then we can obtain the spacelike stretched warped $AdS_3$ black hole solution which is represented by \cite{Anninos:2008fx}
\begin{equation}\label{w-ads-bh}
   ds^2 = -N(r)^2 dt^2 + \frac{\ell^4 dr^2}{4 R(r)^2 N(r)^2} + \ell^2 R(r)^2 \big( d\theta + N^{\theta} (r) dt \big)^2   \,,
\end{equation}
where functions constituting the metric are defined by
\begin{eqnarray}
   && R(r)^2 = \frac{r}{4} \Big( 3(\nu^2 - 1) r + (\nu^2 + 3) (r_+ + r_-)
                       - 4\nu \sqrt{r_+ r_- (\nu^2 + 3)} \Big)   \,,   \nonumber    \\
   && N(r)^2 = \frac{\ell^2 (\nu^2 + 3) (r - r_+) (r - r_-)}{4 R(r)^2}   \,,      \label{def-func-wads} \\
   && N^{\theta} (r) = \frac{2\nu r - \sqrt{r_+ r_- (\nu^2 + 3)}}{2 R(r)^2}  \,.     \nonumber   
\end{eqnarray}
Since the parameter condition for the causally regular black hole solution is $0 < \beta^2 < 1$, we take $\nu > 1$ for convenience. 
From (\ref{funcs-1}) and (\ref{funcs-2}) we can express the gauge field $A$ as follows
\begin{equation}
     A = \phi_1 \Big[ (1 - \beta^2 ) dt + \big( \rho + \omega (1 - \beta^2) \big) d\phi \Big]   \,.
\end{equation}
If we change coordinates and parameters following (\ref{beta-zeta}), then the electromagnetic potential can be represented by
\begin{equation}\label{em-pot}
   A = \phi_1 \ell  \sqrt{\frac{3(\nu^2 - 1)}{4}} \bigg\{ \frac{1}{\nu} dt + \frac{2\nu r - \sqrt{r_+ r_- (\nu^2 + 3)}}{2 \nu} d\theta \bigg\}  \,.
\end{equation}
%
The undetermined parameter $\phi_1$ can be determined by considering the magnitude of the electro-magnetic field. 
We can see that $h_0$ is related to $\phi_0$ from (\ref{funcs-1}), (\ref{rels}), and (\ref{funcs-2}), but $\phi_1$ does not affect this relation. 
The sign of $\phi_1$ means the opposite charge or opposite direction of rotation,  so we  can take $\phi_1$ to be arbitrary. 
It means that the rotation of this black hole can be generated 
by the angular part $A_{\phi}$  from the electromagnetic potential (\ref{em-pot}). This angular part $A_{\phi}$ is also related to (\ref{c-conds}), so it makes a warping factor term $R(\rho)$. 
With the appropriate boundary conditions \cite{Compere:2008cv,Compere:2009zj,Blagojevic:2009ek,Henneaux:2011hv}, asymptotic isometries 
$U(1)_L \times SL(2, \mathbb{R})_R$  can be reduced to Virasoro plus $U(1)$ Kac-Moody algebra at the boundary, suggesting holographically dual warped CFT \cite{Detournay:2012pc}.  
%

\setcounter{equation}{0}
\section{Thermodynamic properties of black holes}
\label{sec-4}
%
In this section, we investigate thermodynamic properties of the warped $AdS_3$ black hole in MMG theory. 
In order to find these properties, we need to find the mass, angular momentum, and entropy of the black hole. These physical quantities can be defined by the Wald formalism \cite{Lee:1990nz,Wald:1993nt}. In \cite{Nam:2016pfp} we have investigated the Wald formalism for the sake of the calculation of the entropy, mass, and angular momentum of black holes with the first order formalism. Here we briefly survey the Wald formalism and definitions of entropy, mass, and angular momentum of black holes. 

If there exists a black hole solution with a local symmetry generated by a Killing vector $\xi$, then the entropy of the black hole is defined on a bifurcation surface, and the corresponding mass and angular momentum are defined well at spatial infinity.  In order to find these definitions we first consider a diffeomorphism invariant theory described by a Lagrangian $d$-form ${\mathscr L}$, where $d$ indicates the space-time dimensions. The variation of a Lagrangian is induced by a field variation
\begin{equation}\label{Lag-var}
     \delta {\mathscr L} = {\mathcal E}_{\psi} \delta \psi + d \Theta (\psi, \delta \psi) \,,
\end{equation}
where $\psi$ describes dynamical fields collectively. 
${\mathcal E}_{\psi} = 0$ means equations of motion constructed by fields variation $\delta \psi$ and $\Theta$ is a $(d-1)$-form ``symplectic 
potential" constructed by dynamical fields $\psi$ and variations of them. 

Let us consider a vector field $\xi$ on a space-time manifold and variations of fields $\psi$ induced by a diffeomorphism generated by this vector,
\begin{equation}
    \delta_{\xi} \psi = \pounds_{\xi} \psi \,.
\end{equation}
Since we are considering a diffeomorphism invariant theory, the variation of the Lagrangian can be represented by the Lie derivative of the Lagrangian under this variation,
\begin{equation}
     \delta_{\xi} {\mathscr L} = \pounds_{\xi} {\mathscr L} = d i_{\xi} {\mathscr L}  \,.
\end{equation}
The above formula means that the vector fields $\xi$ on a space-time generate infinitesimal local symmetries. Applying this formula into (\ref{Lag-var}), we can define 
a closed $(d-1)$-form Noether current, 
\begin{equation}\label{Noe-J-1}
     J_{\xi} = \Theta(\psi \,, \pounds_{\xi} \psi) - i_{\xi} {\mathscr L}   \,,
\end{equation}
under on-shell conditions. So, this current can be described by an exact $(d-2)$-form,
\begin{equation}\label{Noe-J-2}
      J_{\xi} = dQ_{\xi}  \,,
\end{equation}
where $Q_{\xi}$ is constructed from fields and their derivatives.  In diffeomorphism covariant framework \cite{Lee:1990nz,Wald:1993nt}, the phase space is given by the projection of the field configuration space composed of the solutions of field equations with a corresponding symplectic form. The variation $\delta_{\xi} \psi$ under on-shell conditions describes the flow vector corresponding to the one-parameter family of diffeomorphisms generated by $\xi$. Therefore, the variation of the Hamiltonian conjugated to $\xi$ is prescribed by the symplectic form which is defined by 
\begin{equation}
    \delta H_{\xi} = \int_{\Sigma} \omega (\psi, \delta \psi, \pounds_{\xi} \psi) \,,
\end{equation}
where the right-hand side is the symplectic form which is defined by the integration of the symplectic current $\omega$ on the Cauchy surface $\Sigma$. If we take $\xi$ as a symmetry vector field of all dynamical fields, i.e., $\pounds_{\xi} \psi = 0$, and their variation $\delta \psi$ satisfies the linearized equation, then the symplectic current is given by 
\begin{equation}
    \omega (\psi, \delta \psi, \pounds_{\xi} \psi) = \delta \Theta (\psi, \pounds_{\xi} \psi) - \pounds_{\xi} \Theta (\psi, \delta \psi) \,.
\end{equation}
By using the variations of the Noether current (\ref{Noe-J-1}), (\ref{Noe-J-2}), and Lagrangian (\ref{Lag-var}), we can find the variation of the Hamiltonian as follows:
\begin{equation}
    \delta H_{\xi} = \int_{\Sigma} d \delta Q_{\xi} - d i_{\xi} \Theta = \oint_{\partial \Sigma} \delta Q_{\xi} - i_{\xi} \Theta  \,,
\end{equation}
where the second integration should be performed on the boundary of the Cauchy surface $\Sigma$. Since $\xi$ generate a symmetry of solutions of fields $\psi$, 
i.e., $\pounds_{\xi} \psi = 0$, the symplectic current vanishes. 
Therefore $\delta H_{\xi} = 0$ gives us a boundary relation
\begin{equation}\label{vari-id-1}
    0 = \oint_{\partial \Sigma} \delta Q_{\xi} - i_{\xi} \Theta \,.
\end{equation}
If we consider a stationary black hole solution with a Killing vector $\xi$ which vanishes on bifurcation surface $\mathcal H$, then the above integration should be performed 
at interior boundary $\mathcal H$, i.e. bifurcation surface, and at its outer boundary, i.e. spatial infinity of the Cauchy surface $\Sigma$. Therefore the above variational form identity (\ref{vari-id-1}) can be represented by
\begin{equation}\label{vari-id-2}
      \int_{\mathcal H} \delta Q_{\xi} - i_{\xi} \Theta = \int_{\infty} \delta Q_{\xi} - i_{\xi} \Theta \,.
\end{equation}

If we assume that the Killing vector $\xi$ specifies time translation and axial rotation with an angular velocity $\Omega_{\mathcal H}$, i.e., 
\begin{equation}
     \xi = \frac{\partial}{\partial t} + \Omega_{\mathcal H} \frac{\partial}{\partial \phi} \,,
\end{equation}
then (\ref{vari-id-2}) can be suitably in comparison with the black hole thermodynamics 
\begin{equation}
    T_H \delta {\mathcal S}_{BH} = \delta {\mathcal M}- \Omega_{\mathcal H} \delta {\mathcal J} \,,
\end{equation}
where the Hawking temperature is given by $T_H=\kappa_S /2\pi$ with the surface gravity $\kappa_S$. Following the Wald formalism we can define the mass and angular momentum of a black hole as follows \cite{Nam:2016pfp}:
\begin{equation}\label{def-MJ}
              \delta {\mathcal M} = - \frac{1}{8\pi G} \int_{\infty} \delta \chi_{\xi} \Big[ \frac{\partial}{\partial t} \Big] \,,   ~~
                  \delta {\mathcal J} = \frac{1}{8\pi G} \int_{\infty} \delta \chi_{\xi} \Big[ \frac{\partial}{\partial \phi} \Big] \,,
\end{equation}
and the entropy of a black hole can also be defined by
\begin{equation}\label{def-Ent}
      \delta {\mathcal S}_{BH} = - \frac{1}{8\pi G} \cdot \frac{2\pi}{\kappa_S} \bigg( \int_{\mathcal H} \delta \chi_{\xi} \Big[ \frac{\partial}{\partial t} \Big] 
                   + \Omega_{\mathcal H} \int_{\mathcal H} \delta \chi_{\xi} \Big[ \frac{\partial}{\partial \phi} \Big] \bigg)  \,,
\end{equation}
where the charge variation form $\delta \chi_{\xi}$ is given by
\begin{equation}
      \delta \chi_{\xi} = \delta Q_{\xi} - i_{\xi} \Theta  \,.
\end{equation}
If we include the gravitational constant $1/8\pi G$ in the Lagrangian from the beginning, we can omit this constant from definitions (\ref{def-MJ}) and (\ref{def-Ent}).
Now we investigate whether this definition for the black hole entropy is correct or not. First, we consider the Lagrangian for the Chern-Simons-like form \cite{Bergshoeff:2014bia}
\begin{equation}\label{CS-Lag}
    {\mathscr L}_{CSL} = \frac{1}{2} g_{rs} a^r \cdot da^s + \frac{1}{6} f_{rst} a^r \cdot (a^s \times a^t) \,.
\end{equation}
This Lagrangian form is introduced to describe diverse gravity theories, i.e., TMG, NMG, MMG, etc., which may include the Chern-Simons term and some auxiliary fields. The notation $a^r$ means a collection of Lorentz vector valued 1-forms $a^{ra}_{\mu} dx^{\mu}$, where $r$ is a ``flavor" index running $1 \cdots N$. 
In the MMG theory case, flavor $N$ represents fields of this theory, i.e., dreibein $e$, connection 1-form $\omega$, and auxiliary field $h$. 
$g_{rs}$ and $f_{rst}$ represent metric and coupling constants on the flavor space, respectively. To find the charge variation form we first consider the variation of this Lagrangian form which is given by
\begin{equation}
   \delta {\mathscr L}_{CSL} = \delta a^r \cdot \bigg( g_{rs} da^s + \frac{1}{2} f_{rst} a^s \times a^t \bigg) + d \bigg( \frac{1}{2} g_{rs} \delta a^r \cdot a^s \bigg)   \,,
\end{equation}
where the first term gives equations of motion and we can read the symplectic potential from the second term 
\begin{equation}
    \Theta (a, \delta a) = \frac{1}{2} g_{rs} \delta a^r \cdot a^s \,.
\end{equation}
Following the Wald formalism we can obtain the Noether charge 
\begin{equation}
   Q_{\xi} = \frac{1}{2} g_{rs} i_{\xi} a^r \cdot a^s  \,,
\end{equation}
with the on-shell condition.
In order to find the charge variation form we consider the variation of the Noether charge and interior product of the symplectic potential. Then the charge variation form for the Chern-Simons-like Lagrangian (\ref{CS-Lag}) is given by
\begin{equation}\label{ch-CS}
     \delta \chi_{\xi} = g_{rs} i_{\xi} a^r \cdot \delta a^s \,.
\end{equation}
Because the Killing vector $\xi$ vanishes on the bifurcation surface $\mathcal H$, the interior product of the symplectic potential with this vector should be 
vanished, i.e., $i_{\xi} \Theta = 0$. Applying this condition to the variation of the Noether charge, then the variation of this charge is equal to the charge 
variation form (\ref{ch-CS}). Therefore, we can define the entropy of a black hole (\ref{def-Ent}) with the charge variation form in the Chern-Simons-like gravity theories.

The calculation for BTZ black hole entropy in TMG, including gravitational Chern-Simons term, has been performed in \cite{Tachikawa:2006sz}. In that paper, the charge for the black hole entropy has been given by $Q'_{\xi}=Q_{\xi} - C_{\xi}$ where $\delta C_{\xi} = i_{\xi} \Theta + \Sigma_{\xi}$ with a choice $\Sigma_{\xi} = 0$. So, the variation $\delta Q'_{\xi}$ is the same 
definition with $\delta \chi_{\xi}$.

Now we are able to calculate the mass, angular momentum, and entropy of a black hole with definitions (\ref{def-MJ}) and (\ref{def-Ent}).
To find these physical quantities in MMG theory with Maxwell and electromagnetic Chern-Simons terms, we consider the symplectic potential $\Theta$ coming from 
the total derivative term of the variation of the Lagrangian ${\mathscr L}_{tot}$. The symplectic potential from (\ref{var-Lag-t}) is given by 
\begin{equation}\label{sym-pot}
   \Theta = - \sigma \delta \omega \cdot e + \frac{1}{2 \mu} \delta \omega \cdot \omega + \delta e \cdot h -  \delta A \wedge {\mathcal F}   \,.
\end{equation}
Considering the variation of the Lagrangian for a vector field $\xi$, the Noether current is given by 
\begin{equation}\label{N-cur}
      J_{\xi} = \Theta (e, \omega, h, \pounds_{\xi} e, \pounds_{\xi} \omega, \pounds_{\xi} h) - i_{\xi} {\mathscr L} \,.
\end{equation} 
The Noether current is closed when equations of motion are satisfied, so it can be represented by an exact form 
\begin{equation}\label{J-dQ}
          J_{\xi} = dQ_{\xi}  \,. 
\end{equation}          
The symplectic potential (\ref{sym-pot}) can be divided into two parts
\begin{eqnarray}
    && \Theta^{grav} (e, \omega, h, \delta e, \delta \omega, \delta h ) =      
                   - \sigma \delta \omega \cdot e + \frac{1}{2 \mu} \delta \omega \cdot \omega + \delta e \cdot h \,,             \label{sym-grav} \\
   && \Theta^{em} (A, \delta A) = -  \delta A \wedge {\mathcal F}  \,.    \label{sym-em}
\end{eqnarray}   
Combining two definitions (\ref{N-cur}) and (\ref{J-dQ}) we can find conserved charges for gravitational and electromagnetic interactions, respectively,
\begin{eqnarray}
        && Q^{grav}_{\xi} = - \sigma i_{\xi} \omega \cdot e + \frac{1}{2 \mu} i_{\xi} \omega \cdot \omega + i_{\xi} e \cdot h \,,     \\
        && Q^{em}_{\xi} = - i_{\xi} A \cdot {\mathcal F} \,.
\end{eqnarray}
When we calculate the Noether current (\ref{N-cur}), we should consider the Lagrangian ${\mathscr L}$ as two parts. One is ${\mathscr L}^{grav}$, which is related to the gravitational interaction, i.e., terms including $e$, $\omega$, and $h$, and the other is ${\mathscr L}^{em}$, which is related to the gauge field $A$ or $F$.
Changing connection 1-forms $\omega$ into the shifted ones $\Omega = \omega + \alpha h$ with a condition $e \cdot h = 0$, we can obtain 
\begin{eqnarray}
   && \Theta^{grav} = - \sigma \delta \Omega \cdot e + \frac{1}{2 \mu} \delta \Omega \cdot \Omega + (1+\sigma\alpha) \delta e \cdot h      
                      - \frac{\alpha}{2 \mu} \Big( \delta \Omega \cdot h + \delta h \cdot \Omega - \alpha \delta h \cdot h \Big) \,,    \\
   && Q^{grav}_{\xi} = - \sigma i_{\xi} \Omega \cdot e + \frac{1}{2 \mu} i_{\xi} \Omega \cdot \Omega + (1+\sigma\alpha) i_{\xi} e \cdot h      
                      - \frac{\alpha}{2 \mu} \Big( i_{\xi} \Omega \cdot h + i_{\xi} h \cdot \Omega - \alpha i_{\xi} h \cdot h \Big)    \,.
\end{eqnarray}
By using the above symplectic potential and conserved charge, we can find the charge variation form for the gravitational interaction as follows:
\begin{eqnarray}
     \delta \chi^{grav}_{\xi} &=& \delta Q^{grav}_{\xi} - i_{\xi} \Theta^{grav}      \nonumber   \\
         &=& - \sigma \big( i_{\xi} \Omega \cdot \delta e + i_{\xi} e \cdot \delta \Omega \big) + \frac{1}{\mu} i_{\xi} \Omega \cdot \delta \Omega    
                 + (1+\sigma\alpha) \big( i_{\xi} e \cdot \delta h + i_{\xi} h \cdot  \delta e \big)      \nonumber   \\
           && - \frac{\alpha}{\mu} \big( i_{\xi} \Omega \cdot \delta h + i_{\xi} h \cdot \delta \Omega - \alpha i_{\xi} h \cdot \delta h \big)  \,,      \label{var-grav}
\end{eqnarray}
and for the electromagnetic interaction
\begin{equation}\label{var-em}
    \delta \chi^{em}_{\xi} = \delta Q^{em}_{\xi} - i_{\xi} \Theta^{em} = - i_{\xi} A \cdot \delta {\mathcal F} - i_{\xi} {\mathcal F} \cdot \delta A \,.       
\end{equation}
When we consider the thermodynamic relationship, the charge variation form $\delta \chi^{em}_{\xi}$ can be defined by
\begin{equation}
      \Phi_{\mathcal H} \delta Q = \frac{1}{2\pi} \int_{\mathcal H} \delta \chi^{em}_{\xi}  \,,
\end{equation}
where $2\pi$ comes from the solid angle in three-dimensions and it depends on the definition of charge. 
As an example, if we consider only electric potential terms, then $-i_{\xi} A$ part of (\ref{var-em}) describes an electric potential and the integration of $\delta {\mathcal F}$ on the bifurcation surface ${\mathcal H}$ corresponds with the variation of the electric charge.
In order to calculate these charge variation forms we first consider the dreibein from the metric of the black hole (\ref{w-ads-bh})
\begin{equation}\label{drei-wads}
    e^0 = N dt \,,  \quad  e^1 = \frac{\ell^2}{2RN} dr \,, \quad  e^2 = \ell R (d\theta + N^{\theta} dt)  \,.
\end{equation} 
We can find connection 1-forms by applying the dreibein into the torsion free condition (\ref{cmg-eq-1}), 
\begin{eqnarray}
    && \Omega^0 = \frac{R^2 N^{\theta\prime}}{\ell} e^0 + \frac{2NR'}{\ell^2} e^2  \,,       \nonumber   \\
    && \Omega^1 = \frac{R^2 N^{\theta\prime}}{\ell} e^1  \,,      \label{con-Om-sol}    \\
    && \Omega^2 = - \frac{R^2 N^{\theta\prime}}{\ell} e^2 + \frac{2RN'}{\ell^2} e^0  \,,      \nonumber 
\end{eqnarray}
With these connection 1-forms we can find curvature 2-forms by using the definition
\begin{equation}
      R(\Omega) = d\Omega + \frac{1}{2} \Omega \times \Omega \,.
\end{equation}
Then components of curvature 2-forms are given by
\begin{eqnarray}
   && R^0 = \bigg( \frac{\nu^2}{\ell^2} + F(r) \bigg) e^1 \wedge e^2 + G(r) e^1 \wedge e^0  \,,    \nonumber    \\
   && R^1 = - \frac{\nu^2}{\ell^2} e^2 \wedge e^0  \,,           \label{Curv-sol}       \\
   && R^2 = \bigg( - \frac{\nu^2}{\ell^2} + \frac{3(\nu^2 -1)}{\ell^2} + F(r) \bigg) e^0 \wedge e^1 - G(r) e^1 \wedge e^2 \,,      \nonumber 
\end{eqnarray}
where functions $F(r)$ and $G(r)$ are simply represented by
\begin{equation}\label{con-G}
      F(r) = \frac{3(\nu^2 - 1)}{\ell^2} \frac{N^2}{\ell^2} \,,~~~  G(r) = - \frac{3(\nu^2 - 1)}{\ell^2} \frac{RNN^{\theta}}{\ell}  \,.
\end{equation}
Applying the above curvature 2-forms into Eq.(\ref{cmg-eq-2}) with ${\mathcal T}_a = 0$, we can also find auxiliary fields
\begin{eqnarray}
   && h^0 = \frac{1}{2\tilde{\mu}} \bigg( \frac{\nu^2}{\ell^2} - \frac{3(\nu^2 - 1)}{\ell^2} - \alpha \Lambda_0 - 2 F \bigg) e^0 - \frac{1}{\tilde{\mu}} G e^2 \,,     \nonumber   \\
   && h^1 = \frac{1}{2\tilde{\mu}} \bigg( \frac{\nu^2}{\ell^2} - \frac{3(\nu^2 - 1)}{\ell^2} - \alpha \Lambda_0 \bigg) e^1   \,,      \label{Aux-sol}    \\
   && h^2 = \frac{1}{2\tilde{\mu}} \bigg( \frac{\nu^2}{\ell^2} + \frac{3(\nu^2 - 1)}{\ell^2} - \alpha \Lambda_0 + 2 F \bigg) e^2 + \frac{1}{\tilde{\mu}} G e^0 \,.    \nonumber 
\end{eqnarray}
The electromagnetic field can be extracted from (\ref{em-pot}) with $\zeta = \mu_E =  2\nu/\ell$,
\begin{equation}\label{em-pot-2}
    {\mathcal F} = - \frac{\nu}{\ell} A       
               = - \phi_1 \sqrt{\frac{3(\nu^2 - 1)}{4}} \bigg\{dt + \bigg( \frac{2\nu r - \sqrt{r_+ r_- (\nu^2 + 3)}}{2} \bigg) d\theta \bigg\}    \,.     
\end{equation} 
The mass and angular momentum of black holes can be calculated from (\ref{var-grav}), and the electric charge of black holes can also be obtained from (\ref{var-em}).  In order to find this charge variation form we need to calculate non-vanishing interior products by the Killing vector $\xi$ and angle $\theta$ parts of the variations of dreibein, connection 1-forms, auxiliary fields, and electro-magnetic fields. All these calculations are represented in Appendix \ref{app-B}. Then the charge variation form for the black hole mass is given by
\begin{eqnarray}
   \delta \chi^{grav}_{\xi} \Big[ \frac{\partial}{\partial t} \Big]  
          &=& - \frac{\nu (\nu^2 + 3)}{2 \mu} \bigg\{ 1 - \frac{4 \nu \alpha}{\tilde{\mu} \ell}  
               + \frac{\alpha^2}{{\tilde{\mu}}^2} \frac{3(\nu^2 - 1)}{\ell^2} \bigg\}     
                 \bigg( \delta(r_+ + r_-) - \frac{1}{\nu} \delta \sqrt{r_+ r_- (\nu^2 + 3)} \bigg) d\theta   \nonumber   \\ 
      && - \bigg[ \frac{3(\nu^2-1) \ell}{4} \bigg( \sigma-\frac{\nu}{\mu \ell} \bigg) + \frac{\nu^3}{2\mu}      
               - \frac{\nu \ell^2}{2\mu(1+\sigma\alpha)} \bigg( \frac{\nu^2}{\ell^2} + \frac{3(\nu^2 - 1)}{\ell^2} - \alpha \Lambda_0 \bigg)    \nonumber   \\
      && - \frac{\alpha}{\mu} \cdot \frac{\ell}{2 \tilde{\mu}} \bigg( \frac{\nu^2}{\ell^2} + \frac{3(\nu^2 - 1)}{\ell^2} - \alpha \Lambda_0 \bigg)    \nonumber   \\
      &&  \cdot \bigg\{ \frac{\nu \ell}{2} \cdot \frac{\alpha}{2 \tilde{\mu}} \bigg( \frac{\nu^2}{\ell^2}     
              + \frac{3(\nu^2 - 1)}{\ell^2} - \alpha \Lambda_0 \bigg)           
              - \frac{3(\nu^2 - 1)}{4} \bigg\}  \bigg] \delta {\mathcal C} d\theta \,,          \label{charge-var-mass-1}
\end{eqnarray}
where
\begin{equation}
    \delta{\mathcal C} = \frac{(\nu^2 + 3)}{3(\nu^2 - 1)} \delta(r_+ + r_-) - \frac{4\nu}{3(\nu^2 - 1)} \delta \sqrt{r_+ r_- (\nu^2 + 3)}   \,.
\end{equation}
Then we can obtain black hole mass form by using the first definition of (\ref{def-MJ})
\begin{eqnarray}
     {\mathcal M} &=& \frac{\nu(\nu^2 + 3)}{8 G \mu\ell} \bigg\{ \bigg(1 -  \frac{2\nu}{\tilde{\mu} \ell} \alpha \bigg)^2 
                 - (\nu^2 + 3) \frac{\alpha^2}{\tilde{\mu}^2 \ell^2} \bigg\}  \cdot \bigg\{( r_+ + r_- ) - \frac{1}{\nu} \sqrt{r_+ r_- (\nu^2 + 3)} \bigg\}   \nonumber       \\
          && + \frac{1}{4G\ell} \bigg[ \frac{3(\nu^2-1) \ell}{4} \bigg( \sigma-\frac{\nu}{\mu \ell} \bigg) + \frac{\nu^3}{2\mu}       
                 - \frac{\nu \ell^2}{2\mu(1+\sigma\alpha)} \bigg( \frac{\nu^2}{\ell^2} + \frac{3(\nu^2 - 1)}{\ell^2} - \alpha \Lambda_0 \bigg)      \nonumber       \\
          && - \frac{\alpha}{\mu} \cdot \frac{\ell}{2 \tilde{\mu}} \bigg( \frac{\nu^2}{\ell^2} + \frac{3(\nu^2 - 1)}{\ell^2} - \alpha \Lambda_0 \bigg)     
                  \cdot \bigg\{ \frac{\nu \ell}{2} \cdot \frac{\alpha}{2 \tilde{\mu}} \bigg( \frac{\nu^2}{\ell^2}         
                     + \frac{3(\nu^2 - 1)}{\ell^2} - \alpha \Lambda_0 \bigg) - \frac{3(\nu^2 - 1)}{4} \bigg\}  \bigg]         \nonumber       \\
          && \cdot \bigg\{ \frac{(\nu^2 + 3)}{3(\nu^2-1)} ( r_+ + r_- ) - \frac{4\nu}{3(\nu^2 - 1)} \sqrt{r_+ r_- (\nu^2 + 3)} \bigg\}   \,,    \label{BH-mass-1}
\end{eqnarray}
where we should change the gravitational constant $G$ to $G \ell$ to give the correct mass dimension. As $\alpha$ goes to zero with parameter conditions 
$\sigma = 1$ and $\mu = 3\nu /\ell$, the above result becomes the mass of the warped $AdS_3$ black hole in TMG \cite{Anninos:2008fx,Nam:2016pfp,Bouchareb:2007yx}.

As we have already seen before, Eq.(\ref{cmg-eq-3}) can be reduced to (\ref{ch-eq}) with a condition ${\mathcal T}_a = 0$. So, this equation can be solved with all the results: (\ref{drei-wads}), (\ref{con-Om-sol}) and (\ref{Aux-sol}). More details are explained in Appendix \ref{app-A} Considering Eq.(\ref{con-mu-2}), we can take for a special case
\begin{equation}\label{cond-con}
     \sigma = 1 \,, \quad  \mu(1+\alpha) = \frac{3\nu}{\ell}  \,, \quad  \alpha = 2\nu^2 - 3 \,, \quad  \Lambda_0 = - \frac{1}{\ell^2}  \,,
\end{equation}
which satisfy conditions (\ref{con-mu-1}) and (\ref{con-mu-3}). 
%
%
With these parameters and $\zeta^2=4\nu^2/\ell^2$ from (\ref{beta-zeta}), we can find upper and lower limits, i.e., (\ref{FA}) and (\ref{FB}), 
for $\zeta/\mu$ as follows:
\begin{eqnarray}
   && \mathsf{A} = \frac{2}{9} (1+\alpha) = \frac{4}{9} (\nu^2 - 1)  \,,  \\
   && \mathsf{B} = \frac{2(1+\alpha)(5\alpha+13)}{13\alpha+21} = \frac{4(\nu^2 - 1)(5\nu^2 - 1)}{13\nu^2 - 9} \,.
\end{eqnarray}
So \textsf{B} is larger than \textsf{A}, and the ratio,
\begin{equation}
    \frac{\zeta}{\mu} = \frac{2}{3} (1+\alpha) = \frac{4}{3} (\nu^2 - 1) \,,
\end{equation}
is located between \textsf{A} and \textsf{B}. 
Parameter values (\ref{cond-con}) represent a point in parameter space. In Table \ref{table1}, we can find this point to be in the parameters region
\begin{equation}
    \alpha > 1 \,, ~~ \Lambda_0 < \zeta^2/4\alpha \,, ~~ \mathsf{A} < \zeta/\mu < \mathsf{B} \,.
\end{equation}
These regions can also be re-expressed with $\nu^2$ as follows:
\begin{equation}
    \nu^2 > 2 \,, ~~ \Lambda_0 < \frac{\nu^2}{(2\nu^2 - 3) \ell^2} \,, ~~ \mathsf{A} < \zeta/\mu < \mathsf{B} \,.
\end{equation}
%
%
The second part of (\ref{BH-mass-1}) vanishes with these parameters (\ref{cond-con}). 
So the mass of the warped $AdS_3$ black hole in MMG theory can be reduced as follows:
\begin{equation}\label{BH-mass}
    {\mathcal M} = \frac{(\nu^2 + 3)}{16G \nu^2} \bigg( r_+ + r_- - \frac{1}{\nu} \sqrt{r_+ r_- (\nu^2 + 3)} \bigg)   \,.
\end{equation}
To find the angular momentum of the black hole we need to calculate the charge variation form (\ref{var-grav}) with a Killing vector $\xi = \frac{\partial}{\partial \theta}$. But 
this charge variation form $\delta \chi_{\xi} [ \frac{\partial}{\partial \theta}]$ can be simply represented by a total variation of this charge form 
\begin{eqnarray} 
    \chi^{grav}_{\xi} \Big[ \frac{\partial}{\partial \theta} \Big]
        &=&  \bigg[ \sigma \ell R^4 {N^{\theta}}' - \frac{2}{\mu \ell^2} \bigg(NRR' + \frac{\alpha \ell^2}{2\tilde{\mu}} GR \bigg)^2     \nonumber    \\
          && + \frac{1}{2\mu} \bigg\{ R^3 {N^{\theta}}' + \frac{\alpha \ell}{2 \tilde{\mu}} \bigg( \frac{\nu^2}{\ell^2} + \frac{3(\nu^2 - 1)}{\ell^2}       
                - \alpha \Lambda_0 + 2 F \bigg) R \bigg\}^2    \nonumber    \\
          && + (1+\sigma \alpha) \cdot \frac{\ell^2}{2 \tilde{\mu}} \bigg( \frac{\nu^2}{\ell^2} + \frac{3(\nu^2 - 1)}{\ell^2} 
                - \alpha \Lambda_0 + 2 F \bigg) R^2  \bigg] d\theta \,.     \label{ang-charge}
\end{eqnarray}
As $\alpha$ goes to zero, the above charge form is reduced to the one of TMG theory and then gives us correct angular momentum for the warped $AdS_3$ black hole in TMG with parameter conditions $\sigma = 1$ and $\mu = 3\nu / \ell$ \cite{Anninos:2008fx,Nam:2016pfp}. 
From functions (\ref{def-func-wads}) we can calculate each functions in square bracket. 
The value of the angular momentum should be calculated at spatial infinity, so we expand the above charge form (\ref{ang-charge}) as a function of $r$. 
This charge form includes $r^2$, $r$ and constant terms. 
Other terms including negative power of $r$ vanish as $r$ goes to infinity. These calculations are summarized in Appendix \ref{app-C}  
It can be shown that coefficients of $r^2$ and $r$ vanish by using the result of (\ref{cond-con}). 
So, the rest constant term gives a value associated to the charge form for the angular momentum of the black hole after some tedious calculations
\begin{equation}\label{BH-ang-1}
   \chi^{grav}_{\xi} \Big[ \frac{\partial}{\partial \theta} \Big] 
        = - \frac{\ell \nu (\nu^2 + 3)}{24} \bigg[ \frac{3}{2 \nu^2} \bigg( r_+ + r_-   
           - \frac{1}{\nu} \sqrt{r_+ r_- (\nu^2 + 3)} \bigg)^2 - \frac{\nu^2 + 2}{2} (r_+ - r_-)^2 \bigg] d\theta \,.  
\end{equation}
Using the definition of the angular momentum (\ref{def-MJ}) with $\delta \chi^{grav} [\frac{\partial}{\partial \theta}]$, we can obtain the angular momentum of the warped $AdS_3$
black hole in minimal massive gravity 
\begin{equation}\label{BH-angm}
     {\mathcal J} = - \frac{\ell (\nu^2 + 3)}{64 G \nu} \bigg[ \bigg( r_+ + r_- - \frac{1}{\nu} \sqrt{r_+ r_- (\nu^2 + 3)} \bigg)^2      
           - \frac{\nu^2 (\nu^2 + 2)}{3} (r_+ - r_-)^2 \bigg]  \,.    
\end{equation}
To obtain the above correct result we need to consider the change of coordinates to be dimensionful. So, we change the gravitational constant $G$ to $G \ell$. This change makes 
the angular velocity to be dimensionful with $1/\ell$ and rotational Killing vector with $\ell$.

To find the entropy of the warped $AdS_3$ black hole in this theory, we first consider the calculation for the charge variation form from the definition (\ref{def-Ent}). 
So, calculations should be performed at the bifurcation surface $\mathcal H$ with a Killing vector $\xi = \frac{\partial}{\partial t} + \Omega_{\mathcal H} \frac{\partial}{\partial \theta}$ 
\begin{equation}\label{charge-ent}
      \delta \chi_{\xi}^{ent} = \delta \chi_{\xi} \Big[ \frac{\partial}{\partial t} \Big] 
                   + \Omega_{\mathcal H} \cdot \delta \chi_{\xi} \Big[ \frac{\partial}{\partial \theta} \Big]     \,,
\end{equation}
where $\Omega_{\mathcal H}$ is given by (\ref{ang-velo}).
The charge variation form for the entropy of the warped $AdS_3$ black hole is given by (\ref{charge-J}). 
By using parameter conditions (\ref{cond-con}), functions (\ref{def-func-wads}), (\ref{Funcs}), and (\ref{ang-funcs}) with variations of functions in 
Appendix \ref{app-B}, we can rearrange the charge variation form for the entropy of the black hole. Then we can take the radial coordinate value $r=r_+$ at the event horizon ${\mathcal H}$. 
From all of these substitutions we can rearrange the charge variation form as a formula of $r_+$ and $r_-$ with their variations $\delta r_+$, $\delta r_-$ and $\delta R_+$. 
So, the result is given by
\begin{eqnarray}\label{chag-ent}
    \delta \chi_{\xi}^{ent} &=& - \frac{\ell (\nu^2 + 3)}{4} \frac{r_+ - r_-}{R_+} \bigg[ \delta R_+ + R'_+ \delta r_+     
                - \frac{\nu^2 - 1}{2\nu} \delta r_+             \nonumber       \\
              && - \frac{2(\nu^2 - 1)}{3\nu} R'_+ \bigg( \frac{\sqrt{r_+ r_- (\nu^2 + 3)}}{2 r_+}        
                - \frac{3(\nu^2 - 1)}{4} \frac{r_+}{R_+} \bigg) \delta r_+       \nonumber       \\
       && + \frac{2(\nu^2 - 1)}{3\nu} \bigg\{ R'_+ \delta \sqrt{r_+ r_- (\nu^2 + 3)}      
                + \bigg( \frac{\sqrt{r_+ r_- (\nu^2 + 3)}}{2 r_+}         
                + \frac{3(\nu^2 - 1)}{4} \frac{r_+}{R_+} \bigg) \delta R_+  \bigg\}      \nonumber       \\     
       && + \frac{(\nu^2 - 1)(2\nu^2 - 3)}{3\nu^2} \delta R_+      
               - \frac{(\nu^2 - 1)(2\nu^2 - 3)(\nu^2 + 3)}{12 \nu^2} \frac{r_+ - r_-}{R_+} \delta r_+        \nonumber       \\
              && + \frac{(\nu^2 - 1)(2\nu^2 - 3)}{6\nu^2} \bigg( \frac{\sqrt{r_+ r_- (\nu^2 + 3)}}{2 r_+}        
                    - \frac{3(\nu^2 - 1)}{4} \frac{r_+}{R_+} \bigg) \delta r_+ \bigg] d\theta  \,.             \
\end{eqnarray}
Using the variation form $\delta R$ described by (\ref{R-vari}) at the event horizon $r_+$ of the black hole
\begin{equation*}
    \delta R_+ = \frac{\nu^2 + 3}{8} \frac{r_+}{R_+} (\delta r_+ + \delta r_-) 
                   - \frac{\nu}{2} \frac{r_+}{R_+} \delta \sqrt{r_+ r_- (\nu^2 + 3)} \,,
\end{equation*}
and $2R_+ = 2\nu r_+ - \sqrt{r_+ r_- (\nu^2 + 3)}$ appropriately, we can rearrange the charge variation form with variations $\delta r_+$ and $\delta r_-$.
Summarizing the above terms then we can obtain the charge variation form as follows:
\begin{eqnarray}
     \delta \chi_{\xi}^{ent} = - \frac{\ell (\nu^2 + 3)}{4} \frac{r_+ - r_-}{R_+} \frac{1}{6\nu} \bigg[ (\nu^4 + 2\nu^2 + 3) \delta r_+   
              - (\nu^2 - 1)(\nu^2 + 3) \delta r_- - \frac{3}{\nu} \delta \sqrt{r_+ r_- (\nu^2 + 3)} \bigg]  d\theta \,.     \nonumber  \\
\end{eqnarray}
The entropy of a black hole can be defined by
\begin{equation}
     T_H \delta {\mathcal S}_{BH} = - \frac{1}{8\pi G \ell} \int_{\mathcal H} \delta \chi_{\xi}^{ent} \,,
\end{equation}
where we change the Newton constant $G$ into $G \ell$ to make parameters to be dimensionful. Using the Hawking temperature for this warped $AdS_3$ black hole
\begin{equation}
     T_H = \frac{\nu^2 + 3}{4 \pi \ell} \frac{(r_+ - r_-)}{2\nu r_+ - \sqrt{r_+ r_- (\nu^2 + 3)}} = \frac{(\nu^2 + 3)}{8\pi\ell} \frac{r_+ - r_-}{R_+}  \,,
\end{equation}
we can obtain the entropy of the space-like warped $AdS_3$ black hole in this theory which is given by
\begin{equation}\label{BH-ent}
   {\mathcal S}_{BH} = \frac{\pi \ell}{12 G \nu} \bigg[ (\nu^4 + 2 \nu^2 +3) r_+ - (\nu^2 - 1)(\nu^2 + 3) r_- 
            - \frac{3}{\nu} \sqrt{r_+ r_- (\nu^2 + 3)} \bigg]   \,.
\end{equation}
To calculate the variation for the electric charge part, we need to consider the charge variation form for the electromagnetic interaction (\ref{var-em}). 
By using the gauge field representations (\ref{em-pot}) and (\ref{em-pot-2}), the result for the calculation of $\delta \chi_{\xi}^{em}$ vanishes. 
Therefore, the contribution of $\Phi_{\mathcal H} \delta Q$ for the first law of black hole thermodynamics is disappeared. 
So, the entropy, mass and angular momentum of this black hole satisfy the first law of black hole thermodynamics
\begin{equation}\label{BH-Therm}
    \delta {\mathcal M} = T_H \delta {\mathcal S}_{BH} + \Omega_{\mathcal H} \delta {\mathcal J} \,,
\end{equation}
with the dimensionful angular velocity
\begin{equation}
         \Omega_{\mathcal H} = - \frac{2}{(2\nu r_+ - \sqrt{r_+ r_- (\nu^2 + 3)}) \ell}  \,.
\end{equation}
%
%
%
%
Also we obtain the Smarr relation between the mass, angular momentum, and entropy of the black hole  
\begin{equation}\label{Smarr}
      {\mathcal M} = T_H {\mathcal S}_{BH} + 2 \Omega_{\mathcal H} {\mathcal J}  \,.
\end{equation}
The same relation with (\ref{Smarr}) have been obtained in the case of the warped $AdS_3$ black hole in TMG theory 
with Maxwell and electromagnetic Chern-Simons terms in \cite{Moussa:2008sj}. The definition for the charge variation form (\ref{charge-ent}) leads us to 
the same correct results for the entropy of the BTZ black hole, the space-like warped $AdS_3$ black hole in TMG \cite{Tachikawa:2006sz,Hotta:2008yq}.   
It has been shown that the space-like warped $AdS_3$ black hole can be regarded as the ground state with $U(1)_L \times SL(2, {\mathbb R})_R$ symmetry 
in TMG with $\nu > 1$ \cite{Anninos:2008fx}. This black hole solution is a discrete quotient of warped $AdS_3$ space. So, it is conjectured that quantum TMG with suitable 
warped $AdS_3$ boundary conditions is holographically dual to a 2D boundary CFT with a suitable central charges. Under this conjecture we can define left and right moving energies
\begin{equation}
     E_L = \frac{\pi^2 \ell}{6} c_L T_L^2 \,, \quad  E_R = \frac{\pi^2 \ell}{6} c_R T_R^2  \,,
\end{equation}
in terms of left and right central charges and temperatures. They can be defined by
\begin{eqnarray}
     && T_L = \frac{(\nu^2 + 3)}{8 \pi \ell} \bigg( r_+ + r_- - \frac{1}{\nu} \sqrt{r_+ r_- (\nu^2 + 3)} \bigg) \,,    \nonumber   \\
     && T_R = \frac{(\nu^2 + 3)}{8 \pi \ell} (r_+ - r_-) \,,
\end{eqnarray}
which is obtained by considering a quotient along the isometry $\partial_{\theta}$. 
The black hole entropy can be described by 
\begin{equation}
     {\mathcal S}_{BH} = \frac{\pi^2 \ell}{3} \big( c_L T_L + c_R T_R \big)  \,,
\end{equation}
as follows the Cardy formula. According to the result in \cite{Nam:2010ub}, the mass and angular momentum can be represented by left and right moving energies
\begin{equation} 
   {\mathcal M} = \sqrt{\frac{3\beta^4}{4\beta^2-1}} \sqrt{\frac{2c_L}{3\ell} E_L}  \,, \quad  {\mathcal J} = - \ell (E_L - E_R) \,.
\end{equation} 
Even though we did not obtain the warped $AdS_3$ black hole solution without matter term, we can refer (\ref{w-ads-bh}) as a solution for MMG theory with 
matter term contribution ${\mathcal T}_a = 0$.
So the above conjectural results can also be applied with this $AdS_3$ black hole in MMG theory. 
Therefore mass and angular momentum can be represented with $\beta^2 = (\nu^2+3)/4\nu^2$ in (\ref{beta-zeta}) as follows:
\begin{eqnarray}
     {\mathcal M} &=& \frac{(\nu^2 + 3)}{4\nu} \cdot \frac{\pi}{3} c_L T_L  \,,          \\
     {\mathcal J} &=& - \frac{\ell (\nu^2 + 3)}{64G \nu} \bigg[ \bigg( r_+ + r_- - \frac{1}{\nu} \sqrt{r_+ r_- (\nu^2 + 3)} \bigg)^2    
                  - \frac{c_R}{c_L} (r_+ - r_-)^2 \bigg]  \,.
\end{eqnarray}
Comparing the above formulas with (\ref{BH-mass}) and (\ref{BH-angm}), then we can obtain left and right moving central charges as follows:
\begin{equation}\label{cent-ch}
       c_L = \frac{6 \ell}{G \nu (\nu^2 + 3)}   \,,  \quad  c_R = \frac{2\nu (\nu^2 + 2) \ell}{G(\nu^2 + 3)}  \,.
\end{equation}
It has been investigated that the variation of the bulk action including the Chern-Simons term is related to the gravitational anomaly for the boundary CFT theory \cite{Kraus:2005zm}. The bulk variation, i.e. general coordinate variation or local Lorentz transformation, can be described by a boundary integration with a coefficient of Chern-Simons term. The gravitational anomaly for the CFT can also be described by the same integration with a coefficient related to the difference between left and right moving central charges. In the context of $AdS/CFT$ correspondence, the comparison between two values make a relation $c_L - c_R = 96 \pi \eta$ where $\eta$ means the coefficient of the Chern-Simons term. In \cite{Anninos:2008fx} they have found that the difference between two central charges of TMG in warped $AdS_3$ matches the coefficient of gravitational anomaly
\begin{equation}\label{d-c-s}
       c_L - c_R = - \frac{\ell}{G\nu} \,,
\end{equation}
with $\eta = -1/32\pi G \mu$ and $\mu = 3\nu/\ell$. From (\ref{cent-ch}) the difference between central charges of MMG in warped $AdS_3$ is given by 
\begin{equation}\label{d-c-p}
      c_L - c_R = - \frac{2(\nu^2 - 1) \ell}{G\nu} \,.
\end{equation}
If we use two conditions $\mu = 3\nu / \ell (1+\alpha)$ and $\alpha=2\nu^2 - 3$ from (\ref{cond-con}), then $\eta = - 2(\nu^2 - 1)\ell/96\pi G\nu$ gives us the same value as (\ref{d-c-p}). As $\alpha$ goes to zero, we obtain (\ref{d-c-s}) again. So, we may suppose that the calculation of a general coordinate transformation for the action of MMG theory should be the boundary integration with coefficient  $\eta = - (1+\alpha) \ell/96\pi G\nu$.

\setcounter{equation}{0}
\section{Conclusion}
\label{sec-5}
%
In this paper we have constructed a space-like warped $AdS_3$ black hole solution in MMG theory with Maxwell and electro-magnetic Chern-Simons terms. 
In order to find black hole solution we solve equations of motion (\ref{cmg-eq-1}), (\ref{cmg-eq-2}), and (\ref{cmg-eq-3}) using the first order formalism.
This black hole solution can be interpreted into a space-like warped $AdS_3$ black hole by changing the coordinates system into 
a Schwarzschild one with $\nu > 1$, such as that referred to the case in TMG theory \cite{Moussa:2008sj,Anninos:2008fx}.

According to Wald's procedure with the first order formalism, we can define the mass and angular momentum of a black hole as variational forms 
which are related to the integration of a charge variation form $\delta \chi_{\xi}$ at spatial infinity (\ref{def-MJ})  \cite{Nam:2016pfp}. 
The mass of the spacelike warped $AdS_3$ black hole can be given by (\ref{BH-mass-1}) with a parameter $\alpha$. As $\alpha$ goes to 0 with $\mu = 3\nu/\ell$,  the mass of this black hole (\ref{BH-mass-1}) becomes the mass value of the same black hole in TMG. The angular momentum of this black hole can be represented as an integral of the charge form (\ref{charge-ang-alpha}) at spatial infinity with parameter $\alpha$. In the same manner we can also obtain the same value of the charge form in TMG which is 
calculated in (B.10) of  \cite{Nam:2016pfp} as $\alpha$ approaches to 0. Solving equations of motion (\ref{cmg-eq-3}), we can obtain three parameter conditions 
(\ref{con-mu-1}), (\ref{con-mu-2}), and (\ref{con-mu-3}). We can find special parameter values (\ref{cond-con}) satisfying these parameter conditions. 
With these special parameter values (\ref{cond-con}) we have obtained the mass and angular momentum of this black hole, (\ref{BH-mass}) and (\ref{BH-angm}). 

The entropy of the black hole can also be defined by an integral of the charge variation form (\ref{def-Ent}) at the bifurcation surface $\mathcal H$. It is well-known that the entropy of a black hole can be calculated as a conserved charge. But if we consider a theory including non-tensorial terms like a gravitational Chern-Simons term, the definition for the entropy of a black hole should be modified \cite{Tachikawa:2006sz}. In this paper, we have defined the entropy of a black hole as an integral of the charge variation form (\ref{def-Ent}). It can be identified that the calculation of the entropy for the warped $AdS_3$ black hole using definition (\ref{def-Ent}) leads to the correct entropy value appeared in \cite{Anninos:2008fx}. In this paper, we have calculated the entropy of the warped $AdS_3$ black hole (\ref{BH-ent}) in MMG theory. This result gives us the correct relation for the first law of black hole thermodynamics (\ref{BH-Therm}). 
And we also have obtained the Smarr relation (\ref{Smarr}) which describes finite relations between physical values of black holes.

As $\alpha$ goes to 0, $\beta^2$ approaches to 
\begin{equation}
   \beta^2  \rightarrow  \frac{1}{4} - \frac{3 \Lambda_0}{\zeta^2} \,,
\end{equation}
with (\ref{zm1}), (\ref{zm2}) and $\sigma=1$. So, we can find the region for $\Lambda_0$ from $0 < \beta^2 < 1$ as follows:
\begin{equation}
     - \frac{\zeta^2}{4} < \Lambda_0 < \frac{\zeta^2}{12} \quad \Leftrightarrow \quad  - \frac{\nu^2}{\ell^2} < \Lambda_0 < \frac{\nu^2}{3\ell^2} \,,
\end{equation}
where parameter $\zeta^2$ can be changed by $4\nu^2/\ell^2$ coming from (\ref{beta-zeta}).  
With parameter solutions (\ref{cond-con}) of Eqs.(\ref{con-mu-1}), (\ref{con-mu-2}), and (\ref{con-mu-3}), we have found physical quantities and relations of 
the warped $AdS_3$ black hole solution in MMG theory such as mass, angular momentum, entropy, central charges, and Smarr relation.
All physical quantities in MMG theory should approach those of TMG theory as $\alpha$ goes to $0$. If we consider $\nu> 1$ values only, then we can read that 
$\nu$ becomes $\sqrt{3/2}$ from (\ref{cond-con}) as $\alpha$ approaches 0. Substituting $\nu = \sqrt{3/2}$ into (\ref{BH-mass}), (\ref{BH-angm}), (\ref{BH-ent}), and (\ref{cent-ch}), 
we can obtain physical quantities as follows:
\begin{eqnarray}
    && {\cal M} =  \frac{3}{16 G} \Big( r_+ + r_- - \sqrt{3 r_+ r_-} \Big)  \,,  \nonumber   \\  
    && {\cal J} = - \frac{3 \ell}{64G} \sqrt{\frac{3}{2}} \bigg[ \Big( r_+ + r_- - \sqrt{3 r_+ r_-} \Big)^2 - \frac{7}{4} (r_+ - r_-)^2 \bigg] \,, \nonumber   \\
    && {\cal S}_{BH} = \frac{\pi \ell}{G} \sqrt{\frac{2}{3}} \bigg[ \frac{11}{16} r_+ - \frac{3}{16} r_- - \frac{1}{4} \sqrt{3 r_+ r_-} \bigg]   \,,    \\
    && c_L = \frac{4}{3} \sqrt{\frac{2}{3}} \frac{\ell}{G} \,, \qquad c_R = \frac{7}{3} \sqrt{\frac{2}{3}} \frac{\ell}{G} \,.     \nonumber 
\end{eqnarray}
The above values can also be obtained by substituting $\nu=\sqrt{3/2}$ into the formulas of the mass, angular momentum, entropy, and central charges 
of the warped $AdS_3$ black hole coming from TMG theory in \cite{Anninos:2008fx}.  \\


The authors of \cite{Adami:2017phg} have constructed the quasi-local conserved charge for the Chern-Simons-like theories of gravity \cite{Bergshoeff:2014bia} with first order formalism. 
To find this charge they have used the off-shell Abbott-Deser-Tekin (ADT) method \cite{Deser:2002rt,Deser:2002jk,Kim:2013zha,Kim:2013cor} and the field variation with the Lorentz-Lie derivative 
\begin{equation}
    \delta_{\xi} a^{ra} = {\mathfrak L}_{\xi} a^{ra} - \delta^r_{~\omega} d\tilde{\lambda}^a \,.
\end{equation}
The Lorentz-Lie derivative is defined by 
\begin{equation}
    {\mathfrak L}_{\xi} e^a = \pounds_{\xi} e^a + \lambda^a_{~c} e^c  \,,
\end{equation}
where $\tilde{\lambda}^a = 1/2 \cdot \epsilon^a_{~bc} \lambda^{bc}$ is a generator of the local Lorentz transformation and $\pounds_{\xi}$ describes the Lie derivative for a vector 
$\xi$. The purpose in introducing this derivative is to circumvent the divergence of the connection 1-form $\omega^a$ on the bifurcation surface ${\mathcal H}$, even though the interior product of a Killing vector $\xi$ for the connection becomes finite \cite{Jacobson:2015uqa}. In our approach, the variation of the field variables for a vector field $\zeta$ can be described by
\begin{equation}
      \delta_{\zeta} a^r_i = \{ \varphi_{\rm diff} [\zeta] \,, a^r_i \} = \pounds_{\zeta} a^r_i + \cdots \,,
\end{equation}
where $\cdots$ means the term proportional to equations of motion \cite{Bergshoeff:2014pca,Afshar:2014ffa} and $\varphi_{\rm diff} [\zeta]$ is a generator of a diffeomorphism 
associated to a vector field $\zeta$, which comes from the sum of primary constraints derived by a Hamiltonian analysis \cite{Bergshoeff:2014pca}. 
Therefore, it is enough to adapt the Wald formalism for the diffeomorphism invariant Lagrangian as we derive the charge variation form to define physical quantities, i.e., the entropy, electric charge, mass and angular momentum of a black hole \cite{Nam:2016pfp}.

There has been a conjecture that TMG theory with suitable asymptotically stretched $AdS_3$ boundary conditions $\nu > 1$ is holographically dual to a two-dimensional boundary CFT with appropriate left and right moving central charges \cite{Anninos:2008fx}. Applying this conjecture into this black hole solution, we have found left and right moving central charges in MMG theory, respectively. In \cite{Anninos:2008fx} it has been shown that the difference between left and right moving central charges matches the coefficient of the diffeomorphism anomaly with a relation $c_L - c_R = 96 \pi \eta$. Considering MMG theory, we can find (\ref{d-c-p}) with $\eta = - (1+\alpha) \ell/96\pi G\nu$. So, the difference between central charges in this theory seems to appear as a shift by a parameter $\alpha$ in comparison with the TMG case.

%
\section*{Acknowledgement}
J.D.P was supported by a grant from the Kyung Hee University (KHU-20110060) and Basic Science Research Program 
through the National Research Foundation of Korea (NRF) funded by the Ministry of Education (No.2015R1D1A1A01061177).
%

\appendix 
\section*{\Large \bf Appendices}

\renewcommand{\thesection}{\Alph{section}.}
\renewcommand{\theequation}{A.\arabic{equation}}
\setcounter{equation}{0}

\section{Equations for parameter conditions}
\label{app-A}
%
As we have seen before, Eq.(\ref{cmg-eq-3}) can be reduced by a form (\ref{ch-eq}) under a condition ${\mathcal T}_a = 0$. By using the concrete form of the dreibein (\ref{drei}), shifted connection 1-forms (\ref{con-Om-sol}), and auxiliary fields (\ref{Aux-sol}) for the warped $AdS_3$ black hole, (\ref{w-ads-bh}) with (\ref{def-func-wads}), this equation can be solved component by component. 
Then we have five equations 
\begin{eqnarray}
     - \frac{\alpha}{4\tilde{\mu}} \bigg( \frac{\nu^2}{\ell^2} - \frac{3(\nu^2 - 1)}{\ell^2} - \alpha \Lambda_0 \bigg)  
            \bigg( \frac{\nu^2}{\ell^2} + \frac{3(\nu^2 - 1)}{\ell^2} - \alpha \Lambda_0 + 2F \bigg)      
          + \bigg( \frac{3(\nu^2 - 1)}{\ell^2} + F \bigg) \frac{R^2 {N^{\theta}}'}{\ell}       \nonumber   \\
          + G' \frac{2RN}{\ell^2} + G \frac{2NR'}{\ell^2}         
          + \sigma\mu (1+\sigma\alpha) \bigg( \frac{\nu^2}{\ell^2} - \alpha \Lambda_0 +F \bigg) + \tilde{\mu} \Lambda_0 = 0 \,,       \label{con-A}    
\end{eqnarray}
\begin{eqnarray}
   F \frac{2RN'}{\ell^2} + F' \frac{2RN}{\ell^2} + 3 G \frac{R^2 {N^{\theta}}'}{\ell} 
           - \frac{\alpha}{2\tilde{\mu}} \bigg( \frac{\nu^2}{\ell^2} - \frac{3(\nu^2 - 1)}{\ell^2} - \alpha \Lambda_0 \bigg) G    
          + \sigma \mu (1+\sigma\alpha) G = 0 \,,        \label{con-B}   
\end{eqnarray}
\begin{eqnarray}
     - \frac{\alpha}{4\tilde{\mu}} \bigg( \frac{\nu^2}{\ell^2} + \frac{3(\nu^2 - 1)}{\ell^2} - \alpha \Lambda_0 + 2F \bigg)   
               \bigg( \frac{\nu^2}{\ell^2} - \frac{3(\nu^2 -1)}{\ell^2} - \alpha \Lambda_0 - 2F \bigg)    
          - \frac{\alpha}{\tilde{\mu}} G^2 + \frac{2RN'}{\ell^2} G           \nonumber     \\
          - \frac{2NR'}{\ell^2} G + \frac{R^2 {N^{\theta}}'}{\ell} \bigg( \frac{3(\nu^2 - 1)}{\ell^2} + 2 F \bigg) 
          + \sigma \mu (1+\sigma\alpha) \bigg( \frac{\nu^2}{\ell^2} - \alpha \Lambda_0 \bigg) + \tilde{\mu} \Lambda_0 = 0 \,,   \label{con-C}  
\end{eqnarray}
\begin{eqnarray}
     - \frac{\alpha}{4\tilde{\mu}} \bigg( \frac{\nu^2}{\ell^2} - \frac{3(\nu^2 - 1)}{\ell^2} - \alpha \Lambda_0 - 2F \bigg)  
       \bigg( \frac{\nu^2}{\ell^2} - \frac{3(\nu^2 - 1)}{\ell^2} - \alpha \Lambda_0 \bigg) - G' \frac{2RN}{\ell^2} - G \frac{2RN'}{\ell^2}      \nonumber   \\
    - \frac{R^2 {N^{\theta}}'}{\ell} \bigg( 2 \cdot \frac{3(\nu^2 - 1)}{\ell^2} + 3 F \bigg)        
          + \sigma \mu (1+\sigma\alpha) \bigg( \frac{\nu^2}{\ell^2} - \frac{3(\nu^2 - 1)}{\ell^2} 
            - \alpha \Lambda_0 - F \bigg) + \tilde{\mu} \Lambda_0 = 0 \,,       \label{con-D}    
\end{eqnarray}
\begin{eqnarray}
    \bigg( \frac{3(\nu^2 - 1)}{\ell^2} + F \bigg) \frac{2NR'}{\ell^2} - \frac{\alpha}{2\tilde{\mu}} \bigg( \frac{\nu^2}{\ell^2}      
              - \frac{3(\nu^2 - 1)}{\ell^2} - \alpha \Lambda_0 \bigg) G        
    + \frac{R^2 {N^{\theta}}'}{\ell} G + F' \frac{2RN}{\ell^2}        \nonumber     \\
    + \sigma \mu (1+\sigma\alpha) G = 0  \,.          \label{con-E}
\end{eqnarray}
Subtracting (\ref{con-B}) from (\ref{con-E}) describes a relation 
\begin{equation}\label{con-F}
     R^2 {N^{\theta}}^2 = 1 + \frac{N^2}{\ell^2}  \,.
\end{equation} 
From (\ref{con-G}) functions $F(r)$ and $G(r)$ can be represented by simple form
\begin{equation}\label{con-T}
      F(r) = \frac{3(\nu^2 - 1)}{\ell^2} \frac{N^2}{\ell^2} \,, ~~~ G(r) = - \frac{3(\nu^2 - 1)}{\ell^2} \frac{RNN^{\theta}}{\ell}  \,,
\end{equation}
Using (\ref{con-F}) and (\ref{con-T}), we get two relations 
\begin{equation}
    \frac{\ell}{\tilde{\mu}} FRN^{\theta} + \frac{1}{\tilde{\mu}} GN = 0 \,, ~~~
            - \frac{1}{\tilde{\mu}} FN - \frac{\ell}{\tilde{\mu}} GRN^{\theta} = \frac{1}{\tilde{\mu}} \cdot \frac{3(\nu^2 -1)}{\ell^2} N  \,.
\end{equation}
Now we substitute all these solutions to Eq.(\ref{ch-eq}) which comes from (\ref{cmg-eq-3}), then we have three conditions as follows
\begin{eqnarray}
      - \frac{\alpha}{4\tilde{\mu}} \bigg( \frac{\nu^2}{\ell^2} - \frac{3(\nu^2 - 1)}{\ell^2} - \alpha \Lambda_0 \bigg) \bigg( \frac{\nu^2}{\ell^2} 
                  + \frac{3(\nu^2 - 1)}{\ell^2} - \alpha \Lambda_0 \bigg) - \frac{\nu}{\ell} \cdot \frac{3(\nu^2 - 1)}{\ell^2}        \nonumber  \\
                   + \sigma \mu (1+\sigma\alpha) \bigg( \frac{\nu^2}{\ell^2} - \alpha \Lambda_0 \bigg)     
                   + \tilde{\mu} \Lambda_0 = 0  \,.         \label{con-mu-1}   
\end{eqnarray}
\begin{eqnarray}
    - \frac{3\nu}{\ell} - \frac{\alpha}{2 \tilde{\mu}} \bigg( \frac{\nu^2}{\ell^2} - \frac{3(\nu^2 - 1)}{\ell^2} - \alpha \Lambda_0 \bigg)   
            + \sigma \mu (1+\sigma\alpha) = 0 \,.       \label{con-mu-2}    
\end{eqnarray}
\begin{eqnarray}
    - \frac{\alpha}{4 \tilde{\mu}} \bigg( \frac{\nu^2}{\ell^2} - \frac{3(\nu^2 - 1)}{\ell^2} - \alpha \Lambda_0 \bigg)^2     
                    + \sigma \mu (1+\sigma\alpha) \bigg( \frac{\nu^2}{\ell^2} - \frac{3(\nu^2 - 1)}{\ell^2} - \alpha \Lambda_0 \bigg)         \nonumber   \\
             + \frac{2\nu}{\ell} \cdot \frac{3(\nu^2 - 1)}{\ell^2} + \tilde{\mu} \Lambda_0 = 0 \,.         \label{con-mu-3}
\end{eqnarray}
If we take special parameters, 
\begin{equation}\label{para-condi}
     \sigma = 1 \,, \quad  \mu(1+\alpha) = \frac{3\nu}{\ell}  \,, \quad  \alpha = 2\nu^2 - 3 \,, \quad  \Lambda_0 = - \frac{1}{\ell^2}  \,,
\end{equation}
then the above three equations for parameter conditions (\ref{con-mu-1}), (\ref{con-mu-2}) and (\ref{con-mu-3}) are satisfied by these parameters (\ref{para-condi}).


\renewcommand{\theequation}{B.\arabic{equation}}
\setcounter{equation}{0}

\section{Some calculations for charge variation forms}
\label{app-B}
%
The variation of non-vanishing angle $\theta$ part of all relevant fields for the calculation of the charge variation form $\delta \chi_{\xi}$ are given by
\begin{eqnarray}
   \delta e^2 = \ell \delta R d\theta \,, \quad \delta \Omega^0 = \frac{2}{\ell} \delta \big( NRR' \big) d\theta  \,,     \quad
    \delta \Omega^2 = - \delta \big( R^3 {N^{\theta}}' \big) d\theta \,,       \nonumber    \\
           \delta h^0 = - \frac{\ell}{\tilde{\mu}} \delta (GR) d\theta \,,      \quad
    \delta h^2 = \bigg\{ \frac{\ell}{2\tilde{\mu}} \bigg( \frac{\nu^2}{\ell^2} + \frac{3(\nu^2 - 1)}{\ell^2} - \alpha \Lambda_0 \bigg) \delta R   
            + \frac{\ell}{\tilde{\mu}} \delta(FR) \bigg\} d\theta    \,,      \label{vari-form}    \\   
    \delta {\mathcal F} = - \frac{\nu}{\ell} \delta A      
             = - \phi_1 \bigg( \frac{3(\nu^2 - 1)}{4} \bigg)^{1/2}  
                 \frac{\sqrt{r_+ r_- (\nu^2 + 3)}}{4 r_+ r_-} \big( r_- \delta r_+ + r_+ \delta r_- \big) d\theta  \,.          \nonumber
\end{eqnarray}
Non-vanishing interior products of dreibein and connection 1-form for the Killing vector $\xi=\frac{\partial}{\partial t}$ are 
given by 
\begin{eqnarray}
    && i_{\xi} e^0 = N \,, ~~~ i_{\xi} e^2 = \ell R N^{\theta} \,,  \nonumber    \\
       &&  i_{\xi} \Omega^0 = \frac{R^2 {N^{\theta}}' }{\ell} N + \frac{2NR'}{\ell^2} \ell R N^{\theta} \,,   \label{int-t}    \\
            &&  i_{\xi} \Omega^2 = - \frac{R^2 {N^{\theta}}'}{\ell} \ell R N^{\theta} + \frac{2RN'}{\ell^2} N  \,,     \nonumber     
\end{eqnarray}
Non-vanishing interior products of auxiliary fields $h^a$ and electro-magnetic fields ${\mathcal F}$ for Killing vector $\xi=\frac{\partial}{\partial t}$ are given by
\begin{eqnarray}             
   && i_{\xi} h^0 = \frac{1}{2\tilde{\mu}} \bigg( \frac{\nu^2}{\ell^2} - \frac{3(\nu^2 - 1)}{\ell^2} - \alpha \Lambda_0 - 2 F \bigg) N
            - \frac{\ell}{\tilde{\mu}} GRN^{\theta}  \,,     \nonumber   \\
   && i_{\xi} h^2 = \frac{\ell}{2\tilde{\mu}} \bigg( \frac{\nu^2}{\ell^2} + \frac{3(\nu^2 - 1)}{\ell^2} - \alpha \Lambda_0 + 2 F \bigg) RN^{\theta}
            + \frac{1}{\tilde{\mu}} GN  \,,        \label{int-aux-t}    \\
   && i_{\xi} {\mathcal F} = - \frac{\nu}{\ell} i_{\xi} A = - \phi_1 \sqrt{\frac{3(\nu^2 - 1)}{4}}  \,.       \nonumber
\end{eqnarray} 
Non-vanishing interior products of relevant fields for Killing vector $\xi = \frac{\partial}{\partial \theta}$ are given by
\begin{eqnarray}
   && i_{\xi} e^2 = \ell R \,, \quad i_{\xi} \Omega^0 = \frac{2NR'}{\ell^2} \ell R \,, \nonumber  \\
   && i_{\xi} \Omega^2 = - \frac{R^2 {N^{\theta}}'}{\ell} \ell R \,,  \qquad  i_{\xi} h^0 = - \frac{\ell}{\tilde{\mu}} GR \,,    \nonumber  \\
   && i_{\xi} h^2 = \frac{\ell}{2\tilde{\mu}} \bigg( \frac{\nu^2}{\ell^2} + \frac{3(\nu^2 - 1)}{\ell^2} - \alpha \Lambda_0 + 2 F \bigg) R  \,,     \\
   && i_{\xi} {\mathcal F} = - \frac{\nu}{\ell} i_{\xi} A = - \phi_1 \sqrt{\frac{3(\nu^2 - 1)}{4}} \bigg( \frac{2\nu r - \sqrt{r_+ r_- (\nu^2 + 3)}}{2}  \bigg) \,.   \nonumber 
\end{eqnarray}
In order to calculate the charge variation form for the black hole entropy, we should get non-vanishing interior products with the Killing vector 
$\xi = \frac{\partial}{\partial t} + \Omega_{\mathcal H} \frac{\partial}{\partial \theta}$. The results are as follows:
\begin{eqnarray}
       && i_{\xi} e^0 = N \,,  \quad  i_{\xi} e^2 = \ell R (N^{\theta} + \Omega_{\mathcal H}) \,,  \quad   i_{\xi} {\mathcal F} = 0 \,,     \nonumber   \\
       && i_{\xi} \Omega^0 = \frac{R^2 {N^{\theta}}'}{\ell} N + \frac{2NR'}{\ell^2} \ell R (N^{\theta} + \Omega_{\mathcal H})  \,,    \quad
             i_{\xi} \Omega^2 = - \frac{R^2 {N^{\theta}}'}{\ell} \ell R (\Omega_{\mathcal H} + N^{\theta}) + \frac{2RN'}{\ell^2} N \,,      \nonumber  \\
       &&  i_{\xi} h^0 = \frac{1}{2\tilde{\mu}} \bigg( \frac{\nu^2}{\ell^2} - \frac{3(\nu^2 - 1)}{\ell^2} - \alpha \Lambda_0 - 2 F \bigg) N    
                       - \frac{\ell}{\tilde{\mu}} GR (\Omega_{\mathcal H} + N^{\theta})  \,,        \label{forms-ent} \\ 
       && i_{\xi} h^2 = \frac{1}{2\tilde{\mu}} \bigg( \frac{\nu^2}{\ell^2} + \frac{3(\nu^2 - 1)}{\ell^2} 
                       - \alpha \Lambda_0 + 2 F \bigg) \ell R (\Omega_{\mathcal H} + N^{\theta})    
                       + \frac{1}{\tilde{\mu}} GN  \,,   \nonumber
\end{eqnarray}
where the angular velocity of the horizon $\Omega_{\mathcal H}$ is defined by  
\begin{equation}\label{ang-velo}
         \Omega_{\mathcal H} = \left. \frac{d\theta}{dt} \right|_{r_+} = - N^{\theta} (r_+)      
           = - \frac{2}{(2\nu r_+ - \sqrt{r_+ r_- (\nu^2 + 3)})} = - \frac{1}{R_+}  \,.
\end{equation}
As $r$ approaches to the event horizon $r_+$, all terms including $\Omega_{\mathcal H} + N^{\theta}$ vanish.

In order to calculate the charge variation forms more concretely, we first consider the differentiation of all functions of (\ref{def-func-wads}) with respect to $r$. 
The derivatives of these functions are given by
\begin{eqnarray}
   && R' = \frac{R}{2r} + \frac{3(\nu^2 - 1)}{8} \frac{r}{R}   \,,      \nonumber   \\
   && N' = \frac{\ell \sqrt{(\nu^2 + 3)}}{4} \bigg\{ \frac{1}{R} \frac{r^2 - r_+ r_-}{r \sqrt{(r - r_+)(r - r_-)}}    
                 - \frac{3(\nu^2 - 1)}{4} \frac{r \sqrt{(r - r_+)(r - r_-)}}{R^3} \bigg\} \,,            \\
   && {N^{\theta}}' = \frac{\sqrt{r_+ r_- (\nu^2 + 3)}}{2} \frac{1}{r R^2}           
                 - \frac{3(\nu^2 -1)}{8} \frac{r}{R^4} \big(2\nu r - \sqrt{r_+ r_- (\nu^2 + 3)} \big)     \,.     \nonumber 
\end{eqnarray}
From the above formulas and (\ref{def-func-wads}) we can get some relations
\begin{eqnarray}\label{Funcs}    
     && \frac{R^2 N^{\theta \prime}}{\ell} = \frac{\sqrt{r_+ r_- (\nu^2 + 3)}}{2 \ell} \frac{1}{r}         
                         - \frac{3(\nu^2 -1)}{8 \ell} \frac{r}{R^2} \big(2\nu r - \sqrt{r_+ r_- (\nu^2 + 3)} \big)  \,,    \nonumber   \\
     && \frac{2NR'}{\ell^2} = \frac{\sqrt{(\nu^2 + 3)}}{2 \ell} \sqrt{(r - r_+)(r - r_-)}             
                        \cdot \Big( \frac{1}{r} + \frac{3(\nu^2 - 1)}{4} \frac{r}{R^2} \Big)    \,,      \\
     && \frac{2RN'}{\ell^2} = \frac{\sqrt{(\nu^2 + 3)}}{2 \ell} \bigg\{ \frac{2r - r_+ - r_-}{\sqrt{(r - r_+)(r - r_-)}}       
                       - \sqrt{(r - r_+)(r - r_-)} \Big( \frac{1}{r} + \frac{3(\nu^2 - 1)}{4} \frac{r}{R^2} \Big) \bigg\} \,.      \nonumber    
\end{eqnarray}
These functions are helpful to calculate the charge variation form for the mass of the black hole. 
The following functions are helpful to calculate the charge form (\ref{ang-charge}) for the angular momentum of the black hole, 
\begin{eqnarray}\label{ang-funcs}
   && R^4 {N^{\theta}}'  = - \frac{3\nu(\nu^2 -1)}{4}r^2         
              + R \bigg( \frac{R}{2 r} + \frac{3(\nu^2 - 1)}{8} \frac{r}{R} \bigg) \sqrt{r_+ r_- (\nu^2 + 3)}  \,,     \nonumber   \\
   && NRR' = \frac{\ell \sqrt{\nu^2 + 3}}{2} \bigg( \frac{R}{2 r} + \frac{3(\nu^2 - 1)}{8} \frac{r}{R} \bigg)      
              \cdot \sqrt{(r - r_+)(r - r_-)}    \,,                     \\
   && R^3 {N^{\theta}}'  = - \frac{3\nu(\nu^2 -1)}{4} \frac{r^2}{R}             
               + \bigg( \frac{R}{2 r} + \frac{3(\nu^2 - 1)}{8} \frac{r}{R} \bigg) \sqrt{r_+ r_- (\nu^2 + 3)}  \,,  \nonumber   \\
   && GR = - \frac{3(\nu^2 - 1)}{\ell^2} \frac{RNN^{\theta}}{\ell} R \,, \qquad  FR = \frac{3(\nu^2 - 1)}{\ell^2} \frac{N^2}{\ell^2} R \,.      \nonumber 
\end{eqnarray}
Some variations of functions to calculate the charge variation form are represented in detail as follows 
\begin{eqnarray}\label{R-vari}
      \delta R = \frac{(\nu^2 + 3)}{8} \frac{r}{R} (\delta r_+ + \delta r_-)               
                       - \frac{\nu}{4} \frac{r}{R} \frac{\sqrt{r_+ r_- (\nu^2 + 3)}}{r_+ r_-} (r_- \delta r_+ + r_+ \delta r_-)  \,,     
\end{eqnarray}
\begin{eqnarray}
     \delta(NRR') &=& \frac{\ell \sqrt{(\nu^2 + 3)}}{2} \bigg\{ - \Big( \frac{R}{2r} + \frac{3(\nu^2 - 1)}{8} \frac{r}{R} \Big)       
                                \cdot   \frac{\big[ (r - r_-) \delta r_+ + (r - r_+) \delta r_- \big]}{2\sqrt{(r - r_+)(r - r_-)}}         \nonumber   \\  
                        && + \Big( \frac{1}{2r} - \frac{3(\nu^2 - 1)}{8} \frac{r}{R^2} \Big) \sqrt{(r - r_+)(r - r_-)}  \delta R  \bigg\}  \,,   
\end{eqnarray}
\begin{eqnarray}
      \delta(R^3 N^{\theta \prime}) &=& \Big( \frac{R}{2r} + \frac{3(\nu^2 - 1)}{8} \frac{r}{R} \Big)         
                                \cdot \frac{\sqrt{r_+ r_- (\nu^2 + 3)}}{2 r_+ r_-} (r_- \delta r_+ + r_+ \delta r_-)        \nonumber   \\  
                        && + \bigg\{ \frac{\sqrt{r_+ r_- (\nu^2 + 3)}}{2r}         
                                 + \frac{3(\nu^2 - 1)}{8} \frac{r}{R^2} (2\nu r - \sqrt{r_+ r_- (\nu^2 + 3)}) \bigg\} \delta R   \,,       
\end{eqnarray}
\begin{eqnarray}
     \delta(G R) &=& \frac{3(\nu^2 - 1)\sqrt{(\nu^2 + 3)}}{4 \ell^2} \bigg\{ \frac{1}{R^2} (2\nu r - \sqrt{r_+ r_- (\nu^2 + 3)})     
                    \cdot  \sqrt{(r - r_+)(r - r_-)} \delta R       \nonumber   \\  
                    && + \frac{\sqrt{(r - r_+)(r - r_-)}}{R} \frac{\sqrt{r_+ r_- (\nu^2 + 3)}}{2 r_+ r_-} \cdot   ( r_- \delta r_+ + r_+ \delta r_- )      \nonumber   \\  
                    && + \frac{2\nu r - \sqrt{r_+ r_- (\nu^2 + 3)}}{2R \sqrt{(r - r_+)(r - r_-)}} \cdot \big[ (r - r_-) \delta r_+ + (r - r_+) \delta r_- \big] \bigg\}  \,,         
\end{eqnarray}
\begin{eqnarray}
      \delta(F R) = \frac{3(\nu^2 - 1)(\nu^2 + 3)}{4 \ell^2} \bigg\{ - \frac{(r - r_+)(r - r_-)}{R^2} \delta R     
                       - \frac{1}{R} \big[ (r - r_- ) \delta r_+ + (r - r_+) \delta r_-  \big] \bigg\}  \,.      
\end{eqnarray}
%


\renewcommand{\theequation}{C.\arabic{equation}}
\setcounter{equation}{0}

\section{Charge variation forms for the warped $AdS_3$ black hole in MMG}
\label{app-C}
%
Substituting (\ref{int-t}) and (\ref{int-aux-t}) into the charge variation form for the gravitational interaction (\ref{var-grav}), we can obtain the charge variation form for the black hole mass as follows
\begin{eqnarray}
    && \delta \chi^{grav}_{\xi} \Big[ \frac{\partial}{\partial t} \Big]
             = - \sigma \bigg\{ \bigg( - \frac{R^2 N^{\theta \prime}}{\ell} \ell R N^{\theta} + \frac{2RN'}{\ell^2} N \bigg) \ell \delta R    
               - \frac{2}{\ell} N \delta (NRR') - \ell R N^{\theta} \delta (R^3 {N^{\theta}}' ) \bigg\} d\theta     \nonumber   \\
        && \quad - \frac{1}{\mu} \bigg\{ \bigg( \frac{R^2 {N^{\theta}}' }{\ell} N + \frac{2NR'}{\ell^2} \ell R N^{\theta} \bigg) \frac{2}{\ell} \delta(NRR')    
               + \bigg( - \frac{R^2{N^{\theta}}' }{\ell} \ell R N^{\theta} + \frac{2RN'}{\ell^2} N \bigg) \delta(R^3 {N^{\theta}}' )  \bigg\} d\theta   \nonumber    \\
        && \quad + (1+\sigma \alpha) \bigg[ \frac{\ell}{\tilde{\mu}} N \delta (GR)       
               + \ell R N^{\theta} \bigg\{ \frac{\ell}{2\tilde{\mu}} \bigg( \frac{\nu^2}{\ell^2} + \frac{3(\nu^2 - 1)}{\ell^2} - \alpha \Lambda_0 \bigg) \delta R
              + \frac{\ell}{\tilde{\mu}} \delta (FR) \bigg\}        \nonumber    \\
        && \quad + \bigg\{ \frac{\ell}{2\tilde{\mu}} \bigg( \frac{\nu^2}{\ell^2} + \frac{3(\nu^2 - 1)}{\ell^2} - \alpha \Lambda_0 + 2F \bigg) RN^{\theta}   
               + \frac{1}{\tilde{\mu}} GN  \bigg\} \ell \delta R \bigg] d\theta       \nonumber    \\
        && \quad - \frac{\alpha}{\mu} \bigg[ \frac{\ell}{\tilde{\mu}}  \bigg( \frac{R^2 {N^{\theta}}' }{\ell} N + \frac{2NR'}{\ell^2} \ell R N^{\theta} \bigg)  \delta (GR)   
                   \nonumber   \\
        && \quad + \bigg( - \frac{R^2 {N^{\theta}}'}{\ell} \ell R N^{\theta} + \frac{2RN'}{\ell^2} N \bigg)      
                 \cdot \bigg\{  \frac{\ell}{2\tilde{\mu}} \bigg( \frac{\nu^2}{\ell^2} 
              + \frac{3(\nu^2 - 1)}{\ell^2} - \alpha \Lambda_0 \bigg) \delta R + \frac{\ell}{\tilde{\mu}} \delta (FR)  \bigg\}     \nonumber      \\
        && \quad - \bigg\{ \frac{1}{2\tilde{\mu}} \bigg( \frac{\nu^2}{\ell^2} - \frac{3(\nu^2 - 1)}{\ell^2} - \alpha \Lambda_0 - 2F \bigg) N 
              - \frac{\ell}{\tilde{\mu}} GRN^{\theta} \bigg\}     
                   \bigg( \frac{2}{\ell} \delta (NRR') + \frac{\alpha \ell}{\tilde{\mu}} \delta (GR) \bigg)     \nonumber     \\
        && \quad - \bigg\{ \frac{\ell}{2\tilde{\mu}} \bigg( \frac{\nu^2}{\ell^2} + \frac{3(\nu^2 - 1)}{\ell^2} - \alpha \Lambda_0 + 2F \bigg) RN^{\theta} 
              + \frac{1}{\tilde{\mu}} GN  \bigg\}   \nonumber     \\
        && \quad \times \bigg\{ \delta (R^3 {N^{\theta}}' ) 
              + \frac{\alpha \ell}{2\tilde{\mu}} \bigg( \frac{\nu^2}{\ell^2} + \frac{3(\nu^2 - 1)}{\ell^2} - \alpha \Lambda_0  \bigg) \delta R   
                 + \frac{\alpha \ell}{\tilde{\mu}} \delta (FR) \bigg\} \bigg] d\theta  \,.         \label{charge-M}
\end{eqnarray}
The above formula can be reexpressed by (\ref{charge-var-mass-1}) with functions  in (\ref{def-func-wads}), some functions and variations in Appendix \ref{app-B} for the warped $AdS$ black hole. Substituting parameter conditions (\ref{para-condi}) into the above result, then the charge variation form becomes (\ref{BH-mass}).
The charge variation form for the angular momentum can be represented by the total variation of the charge form (\ref{ang-charge}). 
Substituting functions (\ref{def-func-wads}) and (\ref{ang-funcs}) in Appendix \ref{app-B} into the charge form (\ref{ang-charge}), we can find the charge form for the angular momentum of the black hole. This charge form is a lengthy and complicated function of $r$. As $r$ goes to infinity, the negative power of $r$ terms of this charge form vanish. So, this charge form can be represented by a function of the $r^2$ order. We now arrange the coefficients of $r$ order by order.
The coefficient of $r^2$ order of this charge form is given by
\begin{eqnarray}
     r^2 \cdot \frac{3(\nu^2 - 1)}{4} \bigg[ - \sigma \nu \ell  - \frac{(\nu^2 + 3)}{2 \mu} \bigg( 1 - \frac{2 \nu}{\tilde{\mu} \ell} \alpha \bigg)^2   
         + \frac{1}{2 \mu} \bigg\{ - \nu + \frac{\alpha \ell}{2 \tilde{\mu}} \bigg( \frac{\nu^2}{\ell^2} + \frac{3(\nu^2 - 1)}{\ell^2} - \alpha \Lambda_0 \bigg)      \nonumber   \\
     + \frac{\alpha}{\tilde{\mu} \ell} (\nu^2 + 3) \bigg\}^2   
     + (1+\sigma\alpha) \bigg\{ \frac{\ell^2}{2 \tilde{\mu}} \bigg( \frac{\nu^2}{\ell^2} 
            + \frac{3(\nu^2 - 1)}{\ell^2} - \alpha \Lambda_0 \bigg) + \frac{(\nu^2 + 3)}{\tilde{\mu}}   \bigg\} \bigg]  \,.   
\end{eqnarray}
The coefficient of $r$ order is given by
\begin{eqnarray}
      r \cdot \frac{3(\nu^2 - 1)}{4} \bigg[ \sigma \ell \sqrt{r_+ r_- (\nu^2 + 3)}     
          - \frac{\nu^2 + 3}{2 \mu} \bigg\{ \frac{2 \nu}{\tilde{\mu} \ell} \alpha \bigg( 1 -  \frac{2 \nu}{\tilde{\mu} \ell} \alpha \bigg)
         \bigg( {\mathcal C} + \frac{1}{\nu} \sqrt{r_+ r_- (\nu^2 + 3)} \bigg)           \nonumber   \\                    
       - (r_+ + r_-) \bigg( 1 -  \frac{2 \nu}{\tilde{\mu} \ell} \alpha \bigg)^2 \bigg\}      
          + \frac{1}{\mu} \bigg( - \nu + \frac{\alpha \ell}{2 \tilde{\mu}} \bigg( \frac{\nu^2}{\ell^2} + \frac{3(\nu^2 - 1)}{\ell^2} - \alpha \Lambda_0 \bigg)
          +\frac{\alpha}{\tilde{\mu} \ell} (\nu^2 + 3) \bigg)    \nonumber  \\
          \cdot \bigg\{ \nu \bigg( {\mathcal C} + \frac{1}{\nu} \sqrt{r_+ r_- (\nu^2 + 3)} \bigg)      
       - \frac{\alpha}{\tilde{\mu} \ell} (\nu^2 + 3) ({\mathcal C} + r_+ + r_-)                              \nonumber   \\      
         + \frac{\mathcal C}{2} \bigg( - \nu + \frac{\alpha \ell}{2 \tilde{\mu}} \bigg( \frac{\nu^2}{\ell^2} + \frac{3(\nu^2 - 1)}{\ell^2} - \alpha \Lambda_0 \bigg)
         + \frac{\alpha}{\tilde{\mu} \ell} (\nu^2 + 3) \bigg)  \bigg\}            \nonumber   \\                                   
       + (1+\sigma\alpha) \bigg\{ \frac{\ell^2}{2\tilde{\mu}} \bigg( \frac{\nu^2}{\ell^2} 
         + \frac{3(\nu^2 - 1)}{\ell^2} - \alpha \Lambda_0 \bigg) {\mathcal C}   
          - \frac{\nu^2 + 3}{\tilde{\mu}} (r_+ + r_-) \bigg\} \bigg]   \,,
\end{eqnarray}
where 
\begin{equation}
     {\mathcal C} = \frac{\nu^2 + 3}{3(\nu^2 - 1)} (r_+ + r_-) - \frac{4\nu}{3(\nu^2 - 1)} \sqrt{r_+ r_- (\nu^2 + 3)} \,.
\end{equation}
The coefficient of $r^0$ order, i.e., the constant term, is given by 
\begin{eqnarray}\label{app-angm}
    \frac{3(\nu^2 - 1)}{4} \bigg[ \sigma \ell \cdot \frac{\mathcal C}{2} \sqrt{r_+ r_- (\nu^2 + 3)}     
          - \frac{\nu^2 + 3}{2 \mu} \bigg\{ r_+ r_- \bigg( 1 - \frac{2 \nu}{\tilde{\mu} \ell} \alpha \bigg)^2 
         - (r_+ + r_- ) \frac{2 \nu}{\tilde{\mu} \ell} \alpha \bigg( 1 - \frac{2 \nu}{\tilde{\mu} \ell} \alpha \bigg)     \nonumber  \\
              \cdot \bigg( {\mathcal C} + \frac{1}{\nu} \sqrt{r_+ r_- (\nu^2 + 3)} \bigg)                                        
         + \frac{1}{4} \bigg( {\mathcal C} - \frac{2\nu}{\tilde{\mu} \ell} \alpha \bigg( 2 {\mathcal C} 
         + \frac{1}{\nu} \sqrt{r_+ r_- (\nu^2 + 3)} \bigg) \bigg)^2 \bigg\}                   \nonumber   \\ 
          + \frac{1}{2 \mu} \bigg\{ \nu \bigg( \frac{\mathcal C}{2} + \frac{1}{\nu} \sqrt{r_+ r_- (\nu^2 + 3)} \bigg)    
                     - \frac{\alpha}{\tilde{\mu} \ell} (\nu^2 + 3) \bigg( \frac{\mathcal C}{2} + r_+ + r_- \bigg)         \nonumber   \\             
           + \frac{\mathcal C}{2} \cdot \frac{\alpha \ell}{2\tilde{\mu}} \bigg( \frac{\nu^2}{\ell^2} 
                         + \frac{3(\nu^2 - 1)}{\ell^2} - \alpha \Lambda_0 \bigg) \bigg\}^2           
               +\frac{1}{\mu} \bigg( -\nu + \frac{\alpha \ell}{2\tilde{\mu}} \bigg( \frac{\nu^2}{\ell^2} + \frac{3(\nu^2 - 1)}{\ell^2} - \alpha \Lambda_0 \bigg)      \nonumber   \\ 
                    + \frac{\alpha}{\tilde{\mu} \ell} (\nu^2 + 3) \bigg)                           
          \cdot \bigg\{ \frac{3{\mathcal C}^2}{8} \bigg( - \nu + \frac{\alpha \ell}{2 \tilde{\mu}} \bigg( \frac{\nu^2}{\ell^2} 
                + \frac{3(\nu^2 - 1)}{\ell^2} - \alpha \Lambda_0 \bigg) + \frac{\alpha}{\tilde{\mu} \ell} (\nu^2 + 3) \bigg)       \nonumber  \\            
                      - \frac{{\mathcal C}^2}{2} \frac{\alpha \ell}{2\tilde{\mu}} \bigg( \frac{\nu^2}{\ell^2}    
                      + \frac{3(\nu^2 - 1)}{\ell^2} - \alpha \Lambda_0 \bigg)   
          + \frac{\alpha}{\tilde{\mu} \ell} (\nu^2 + 3) (r_+ + r_-) \frac{\mathcal C}{2} + \frac{\alpha}{\tilde{\mu} \ell} (\nu^2 + 3) r_+ r_- \bigg\}      \nonumber  \\
               + (1+\sigma\alpha) \frac{\nu^2 +3}{\tilde{\mu}} r_+ r_- \bigg] \,.
\end{eqnarray}
This constant term is related to the angular momentum of the black hole.
Rearranging the above coefficients with parameter conditions (\ref{para-condi}), the coefficients of $r^2$ and $r$ disappear.  Only the constant term survives.
The charge form for the angular momentum of the black hole with $\sigma = 1$ can be represented by
\begin{eqnarray}
    \chi^{grav}_{\xi} \Big[ \frac{\partial}{\partial \theta} \Big] &=&
             \frac{3(\nu^2 - 1)}{4} \bigg[ \frac{\ell}{2} {\mathcal C} \sqrt{r_+ r_- (\nu^2 + 3)}  
              - \frac{\nu^2+3}{2\mu} \bigg( \frac{1}{4} {\mathcal C}^2 + r_+ r_- \bigg)   \nonumber  \\
       && +  \frac{\nu^2}{2\mu} \bigg( \frac{\mathcal C}{2} + \frac{1}{\nu} \sqrt{r_+ r_- (\nu^2 + 3)} \bigg)^2   
             + \frac{3\nu^2}{8 \mu} {\mathcal C}^2 + \frac{(\nu^2 + 3)}{\mu(1+\alpha)} r_+ r_-      \nonumber  \\
       && + \frac{(\nu^2+3)}{2\mu} \bigg\{ 2 \cdot \frac{2\nu \alpha}{\tilde{\mu} \ell} \bigg( 1 - \frac{1}{2} \cdot \frac{2\nu\alpha}{\tilde{\mu} \ell} \bigg) r_+ r_-  
                  \nonumber  \\
       && + (r_+ + r_-) \frac{2\nu \alpha}{\tilde{\mu} \ell} \bigg( 1 - \frac{2\nu\alpha}{\tilde{\mu} \ell} \bigg) \bigg({\mathcal C} 
             + \frac{1}{\nu} \sqrt{r_+ r_- (\nu^2 + 3)} \bigg)           \nonumber      \\
       && + \frac{1}{2} \cdot \frac{2\nu \alpha}{\tilde{\mu} \ell} {\mathcal C} \bigg( 2{\mathcal C} + \frac{1}{\nu} \sqrt{r_+ r_- (\nu^2 + 3)} \bigg)   
             - \frac{1}{4} \bigg( \frac{2\nu \alpha}{\tilde{\mu} \ell} \bigg)^2 \bigg( 2{\mathcal C} + \frac{1}{\nu} \sqrt{r_+ r_- (\nu^2 + 3)} \bigg)^2 \bigg\}  \nonumber   \\
       && + \frac{1}{2\mu} \bigg\{ \bigg( \frac{\alpha}{\tilde{\mu} \ell} (\nu^2 + 3) \bigg)^2 \Big({\mathcal C} + r_+ + r_- \Big)^2      \nonumber  \\
       && + \frac{{\mathcal C}^2}{4} \bigg( \frac{\alpha \ell}{2\tilde{\mu}}\bigg( \frac{\nu^2}{\ell^2} + \frac{3(\nu^2 - 1)}{\ell^2} - \alpha \Lambda_0 \bigg) 
             + \frac{\alpha}{\tilde{\mu} \ell} (\nu^2 + 3) \bigg)^2           \nonumber        \\
       && - 2 \nu \frac{\alpha}{\tilde{\mu} \ell} (\nu^2 + 3) \bigg( \frac{\mathcal C}{2} + \frac{1}{\nu} \sqrt{r_+ r_- (\nu^2 + 3)} \bigg) \Big({\mathcal C} + r_+ + r_- \Big)  
                \nonumber       \\
       && + \nu \bigg( \frac{\alpha \ell}{2\tilde{\mu}}\bigg( \frac{\nu^2}{\ell^2} + \frac{3(\nu^2 - 1)}{\ell^2} - \alpha \Lambda_0 \bigg)
             + \frac{\alpha}{\tilde{\mu} \ell} (\nu^2 + 3) \bigg)       
             \cdot {\mathcal C} \bigg( \frac{\mathcal C}{2} + \frac{1}{\nu} \sqrt{r_+ r_- (\nu^2 + 3)} \bigg)    \nonumber          \\
       && - \frac{\alpha}{\tilde{\mu} \ell} (\nu^2 + 3) \bigg( \frac{\alpha \ell}{2\tilde{\mu}}\bigg( \frac{\nu^2}{\ell^2} + \frac{3(\nu^2 - 1)}{\ell^2} 
             - \alpha \Lambda_0 \bigg)       
             + \frac{\alpha}{\tilde{\mu} \ell} (\nu^2 + 3) \bigg) {\mathcal C} \Big( {\mathcal C} + r_+ + r_- \Big) \bigg\}       \nonumber          \\
       && - \frac{1}{\mu} \bigg( \frac{\alpha \ell}{2\tilde{\mu}}\bigg( \frac{\nu^2}{\ell^2} + \frac{3(\nu^2 - 1)}{\ell^2} - \alpha \Lambda_0 \bigg) 
             + \frac{\alpha}{\tilde{\mu} \ell} (\nu^2 + 3) \bigg) \cdot \frac{3\nu}{8} {\mathcal C}^2   \nonumber     \\
       && + \frac{1}{\mu} \bigg( - \nu + \frac{\alpha \ell}{2 \tilde{\mu}} \bigg( \frac{\nu^2}{\ell^2} + \frac{3(\nu^2 - 1)}{\ell^2} - \alpha \Lambda_0 \bigg) 
             + \frac{\alpha}{\tilde{\mu} \ell} (\nu^2 + 3) \bigg)       \nonumber  \\
       &&   \cdot  \bigg\{ - \frac{{\mathcal C}^2}{8} \cdot \frac{\alpha \ell}{2 \tilde{\mu}} \bigg( \frac{\nu^2}{\ell^2} + \frac{3(\nu^2 - 1)}{\ell^2} 
             - \alpha \Lambda_0 \bigg)    \nonumber  \\
       && + \frac{\alpha}{\tilde{\mu} \ell} (\nu^2+3) \cdot \frac{3 {\mathcal C}^2}{8}    
             + \frac{\alpha}{\tilde{\mu} \ell} (\nu^2+3) (r_+ + r_-) \frac{\mathcal C}{2} \bigg\}      
             + \frac{\alpha}{\tilde{\mu} \ell} (\nu^2+3) r_+ r_- \bigg]  \,.                \label{charge-ang-alpha}
\end{eqnarray}
Substituting parameter conditions (\ref{para-condi}) into the above formula, then the charge form becomes (\ref{BH-ang-1}).  
The charge variation form for the entropy of the black hole can be obtained by adding the charge variation form $\delta \chi_{\xi}^{grav} [ \frac{\partial}{\partial t}]$ 
to the charge variation form for $\delta \chi_{\xi}^{grav} [ \frac{\partial}{\partial \theta}]$ multiplied by angular velocity $\Omega_{\mathcal H}$ (\ref{ang-velo}). 
By using (\ref{vari-form}) and (\ref{forms-ent}), the charge variation form for the entropy of the black hole is given by
\begin{eqnarray}
     \delta \chi_{\xi}^{ent} &=& \delta \chi_{\xi}^{grav} \Big[ \frac{\partial}{\partial t} \Big] 
                              + \Omega_{\mathcal H} \cdot \delta \chi_{\xi}^{grav} \Big[ \frac{\partial}{\partial \theta} \Big]       \nonumber     \\
          &=& - \sigma \bigg\{ \frac{2RN'}{\ell} N \delta R - \frac{2}{\ell} N \delta (NRR') \bigg\} d\theta   
                  + (1+\sigma\alpha) \frac{\ell}{\tilde{\mu}} \bigg\{ N \delta (GR) + GN \delta R       \bigg\} d\theta                \nonumber       \\
          &&   - \frac{1}{\mu} \bigg\{ \frac{2 R^2 {N^{\theta}}'}{\ell^2} N \delta (NRR') + \frac{2RN'}{\ell^2} N \delta (R^3 {N^{\theta}}') \bigg\} d\theta          \nonumber       \\     
          &&   - \frac{\alpha}{\mu} \bigg[ \bigg( \frac{2RN'}{\ell^2} N - \frac{\alpha}{\tilde{\mu}} GN  \bigg) 
                  \cdot \frac{\ell}{2 \tilde{\mu}} \bigg( \frac{\nu^2}{\ell^2} + \frac{3(\nu^2 - 1)}{\ell^2} - \alpha \Lambda_0 \bigg)  \delta R             \nonumber \\  
          &&   + \bigg\{ \frac{1}{\tilde{\mu}} F - \frac{1}{2 \tilde{\mu}} \bigg( \frac{\nu^2}{\ell^2} - \frac{3(\nu^2 - 1)}{\ell^2} 
                  - \alpha \Lambda_0 \bigg) \bigg\}   \frac{2}{\ell} N \delta (NRR')               \nonumber      \\
           &&  - \frac{1}{\tilde{\mu}} GN \delta (R^3 {N^{\theta}}') 
                 + \frac{\ell}{\tilde{\mu}} \bigg( \frac{2RN'}{\ell^2} N - \frac{\alpha}{\tilde{\mu}} GN \bigg) \delta (FR)               \nonumber \\
          &&  + \frac{\ell}{\tilde{\mu}} \bigg\{ \frac{R^2 {N^{\theta}}'}{\ell} N - \frac{\alpha}{2 \tilde{\mu}} \bigg( \frac{\nu^2}{\ell^2} 
                 - \frac{3(\nu^2 - 1)}{\ell^2} - \alpha \Lambda_0 \bigg) N + \frac{\alpha}{\tilde{\mu}} FN \bigg\} \delta (GR) \bigg] d\theta  \,.    \label{charge-J}
\end{eqnarray}
Substituting (\ref{def-func-wads}), (\ref{Funcs}), (\ref{ang-funcs}) and variations of function from below (\ref{R-vari}) into the above formula, 
this charge variation form becomes (\ref{chag-ent}) at the horizon $r=r_+$.


\newpage

\end{document}